\documentclass[sigconf, nonacm]{acmart}
\usepackage{xspace}
\usepackage{soul}
\usepackage{multirow}
\usepackage{booktabs}
\usepackage{tabularx}
\usepackage{array}
\usepackage{float}
\usepackage{placeins}
\AtBeginDocument{%
  }

\newcommand{\systemName}{\textsc{DISARM}\xspace}

\setcopyright{none}

\begin{document}

\title[Disarming the Hidden Puppetmaster]{\systemName-ing the Hidden Puppetmaster: An Interface for Exposing Manipulative Tactics in Large Language Models}


\author{Kynnedy Simone Smith}
\email{kynnedys@andrew.cmu.edu}
\affiliation{%
  \institution{Carnegie Mellon University}
  \city{Pittsburgh}
  \state{PA}
  \country{USA}
}

\author{Kantwon Rogers}
\email{kantwon@media.mit.edu}
\affiliation{%
  \institution{Massachusetts Institute of Technology}
  \city{Cambridge}
  \state{MA}
  \country{USA}
}

\author{Priyal Shrivastava}
\email{priyalshri1@gmail.com}
\affiliation{%
  \institution{Carnegie Mellon University}
  \city{Pittsburgh}
  \state{PA}
  \country{USA}
}

\author{Motahhare Eslami}
\email{meslami@andrew.cmu.edu}
\affiliation{%
  \institution{Carnegie Mellon University}
  \city{Pittsburgh}
  \state{PA}
  \country{USA}
}

\author{Jocelyn Shen}
\email{joceshen@media.mit.edu}
\affiliation{%
  \institution{Massachusetts Institute of Technology}
  \city{Cambridge}
  \state{MA}
  \country{USA}
}

\author{Maarten Sap}
\email{msap@andrew.cmu.edu}
\affiliation{%
  \institution{Carnegie Mellon University}
  \city{Pittsburgh}
  \state{PA}
  \country{USA}
}

\renewcommand{\shortauthors}{Smith et al.}

\begin{abstract}
Large language models (LLMs) can covertly steer users’ beliefs through emotionally manipulative tactics, yet little is known about how interfaces can protect users as such influence unfolds. We investigate whether in-situ disclosures can inoculate users against manipulation during AI conversations. Following a formative study, we developed \systemName, an interface that identifies, explains, and tracks manipulative tactics and hidden incentives. We evaluated our interface in a between-subjects experiment where 488 participants sought advice from an LLM pursuing a harmful hidden objective, with or without disclosures from \systemName. Participants receiving disclosures exhibited 78\% less belief shift toward the agent’s objective and perceived its advice as more biased and less actionable and the agent as less ethical, reliable, and well-intentioned. An inductive analysis of 1,927 user turns identified 16 response and pushback strategies. We contribute evidence that real-time disclosures can reduce manipulative influence, design recommendations for resistance, and a dataset of categorized responses.

\end{abstract}


\ccsdesc[500]{Human-centered computing~Empirical studies in HCI}
\ccsdesc[500]{Human-centered computing~Interactive systems and tools}
\ccsdesc[500]{Human-centered computing~User studies}


\keywords{large language models, AI manipulation, human-AI interaction, persuasive systems, manipulation disclosure, belief change}


\maketitle



\section{Introduction}

People are increasingly turning to large language models (LLMs) for advice and guidance \cite{chatterji_how_nodate}. Although seeking advice can be beneficial, it also exposes people's beliefs and decisions to the values and incentives of the advice provider. This creates particular risks when the provider is an AI system whose objectives and sources of influence may be difficult for users to identify. Models may implicitly steer users through incentives introduced during training, alignment, or system design \cite{kirk_personalisation_2023,weidinger_taxonomy_2022}. Preference-alignment pipelines can also inadvertently reward agreeable or sycophantic responses, reinforcing users' existing views and potentially increasing their dependence on AI advice \cite{cheng_sycophantic_2025,sharma_towards_2024}. In these cases, manipulation may emerge from opaque system-level processes rather than an explicitly malicious instruction, making its source especially difficult for users to recognize.

Manipulative agents can also be created more directly through prompting. When instructed only to be persuasive, LLMs have been shown to employ tactics such as flattery, emotional appeals, and false authority without being told which tactics to use \cite{liu_llm_2025,shi_siren_2025,freitas_emotional_2025}. When given hidden objectives, conversational agents have pushed users toward harmful financial and emotional decisions and produced measurable changes in their beliefs \cite{sabour_human_2025,shen2026puppet}. These findings demonstrate that manipulative influence can arise both implicitly from how AI systems are designed and explicitly from objectives supplied at deployment. Yet, in either case, users may experience the resulting responses simply as personalized or helpful advice.

Despite growing concern about online manipulation, persuasion, and misinformation, relatively little work has examined how to protect users while manipulation is unfolding in a live AI conversation. Existing work on AI manipulation has primarily characterized manipulative behaviors, developed detection datasets and persuasion-safety benchmarks, or measured users' susceptibility after interacting with manipulative agents \cite{contro_chatbotmanip_2025,liu_llm_2025,sabour_human_2025,shi_siren_2025,freitas_emotional_2025}. In adjacent misinformation and online-persuasion domains, HCI research has explored user-facing interventions such as warning labels, explanations, intent disclosures, and debunking \cite{jamieson_flagging_2025-1,kirchner_countering_2020,cemiloglu_explainability_2023}. These approaches offer useful design foundations, but they do not resolve the distinctive challenges of intervening in an adaptive, multi-turn conversation.

In this work, we investigate how chat interfaces can inoculate users against emotionally manipulative persuasion tactics from LLMs. Designing such interfaces requires addressing two central challenges. First, conversational manipulation is dynamic: unlike a static post or interface element, an AI agent can shift its arguments across turns and adapt its responses to information revealed by the user \cite{salvi_conversational_2025,sabour_human_2025}. Moreover, whereas misinformation interventions can provide accurate information, reliable sources, or direct corrections, learning that an agent is manipulating them does not necessarily tell users how to interpret its advice or respond. Second, an intervention must make manipulation sufficiently visible without overwhelming users or encouraging indiscriminate distrust. Explanations can introduce additional cognitive effort, while poorly calibrated warnings may make users overly skeptical of AI-mediated information more broadly \cite{vasconcelos_explanations_2023,sahebi_ai-mediated_2025}. More research is therefore needed to understand how interfaces can expose and explain manipulation in real time while preserving users' ability to engage with AI productively.

To address this gap, we designed and evaluated \systemName, a real-time intervention that helps users recognize, understand, and resist emotionally manipulative persuasion as it unfolds during interactions with AI agents. We developed the intervention through a formative study that informed three core features: (1) turn-by-turn disclosure of manipulation tactics and hidden incentives, (2) language-level highlighting of manipulative content, and (3) a dynamic summary that tracks manipulation across the conversation. Together, these features make manipulative influence visible at multiple levels, from the specific language used in an individual response to the broader objectives and behavioral patterns of the agent. We examine how this transparency shapes users' susceptibility to manipulation, their perceptions of the agent and its advice, and the ways they respond when manipulation is surfaced. We therefore ask:

\noindent\hangindent=2.6em\hangafter=1 \textbf{RQ1}\hspace{0.5em} How does exposing and explaining manipulative tactics and incentives affect users’ belief shift during interactions with a manipulative AI agent?

\noindent\hangindent=2.6em\hangafter=1 \textbf{RQ2}\hspace{0.5em} How does this transparency shape users’ perceptions, evaluations, and trust of AI agents?

\noindent\hangindent=2.6em\hangafter=1 \textbf{RQ3}\hspace{0.5em} How do users respond when manipulative influence is made visible, and what strategies do they employ to interpret, resist, or push back against a manipulative AI agent?

To answer these questions, we conducted a controlled between-subjects online study ($N = 488$) in which participants sought advice from an AI agent pursuing a harmful hidden objective, either with or without manipulation disclosures from \systemName. Participants who used \systemName exhibited 78\% less belief shift toward the agent's objective than participants in the control condition. Insights from \systemName also led participants to perceive the advice as more biased and to evaluate the agent's intentions, ethics, and reliability less favorably. Through an inductive analysis of 1,927 user turns, we further identified 16 strategies through which users engaged with, resisted, and attempted to regulate the manipulative agent. Our contributions are: (1) empirical evidence that real-time explanatory disclosures can reduce manipulative belief influence and recalibrate users' perceptions of AI advice; (2) an analysis of how users respond and push back after manipulation is made visible; (3) a novel dataset of 1,927 user turns categorized by response strategy; and (4) design implications for helping users recognize, resist, regulate, or disengage from manipulative AI interactions.

\section{Background and Related Work}


Manipulation is a form of social influence that shapes another person's beliefs, emotions, or decisions by exploiting vulnerabilities in how they make judgments \cite{fischer_then_2022,noggle_ethics_2018,susser_online_2018,werner_influencer-centered_2024}. Although manipulation overlaps with persuasion, the two are not equivalent. Persuasion broadly seeks to influence attitudes or behavior and can do so through reasons that a person can consciously evaluate. Manipulation instead often operates covertly, obscuring the influencer's intentions or using emotional and cognitive vulnerabilities to bypass deliberative scrutiny \cite{susser_online_2018,noggle_ethics_2018}. For example, emotional framing can influence judgments through affective or peripheral cues rather than the merits of an argument itself \cite{finucane_affect_2000,petty_elaboration_1986,petty_emotion_2015}. The most dangerous yet effective types of manipulation influences a target without them fully realizing how or why their judgment is being shaped. As conversational technologies increasingly participate in users' decision-making, an open question is how these forms of influence manifest when the manipulator is an AI system. In this section, we first examine the concept of AI manipulation and its consequences, then review existing approaches to mitigating this phenomenon, and finally identify the key gaps that our work aims to address.

\subsection{AI Manipulation}

Focusing on the hidden influence quality of manipulation, we define \textit{AI manipulation} as an AI system covertly steering a user's beliefs, emotions, or decisions toward an objective that is hidden from the user \cite{klenk_ethics_2024,susser_online_2018}. AI systems can draw on several forms of influence to steer users towards both implicit and hidden incentives \cite{matz_potential_2024,salvi_conversational_2025}. They can also exhibit sycophancy to validating users in ways that reinforce their existing judgments \cite{cheng_sycophantic_2025}, or employ emotional tactics such as guilt, fear of missing out, and appeals for sympathy to manipulate users as well \cite{freitas_emotional_2025}. Most directly, \citet{sabour_human_2025} found that conversational agents given hidden objectives could steer users toward harmful financial and emotional decisions even without explicit instructions about which manipulation tactics to employ. The consequences of AI manipulation range from poor decisions to severe psychological harms, including AI-associated psychosis and self-harm \cite{augustin_characterizing_2026,hudon_delusional_2025,pierre_youre_2025,brittain_google_2026}.

We must study real-time experiences of these negative influences in order to understand how to best protect users and improve AI systems. However, we must also be careful to choose probes that do not cause long term harm to users. Recent work from \citet{shen2026puppet} has found that agents pursuing hidden incentives can produce short term, measurable changes in users' beliefs during everyday advice conversations. Their work suggests that belief shift can serve as a measurable proximal effect of AI manipulation. Building on this approach, our research contributes a novel exploration of real-time conversational AI manipulation by adapting the structure of their manipulative AI model and using belief shift to evaluate whether a manipulation disclosure interface can intervene with the effects of negative influence. 

\subsection{Combating AI Manipulation}
Existing work on AI manipulation has largely focused on characterizing and measuring the problem rather than intervening during an ongoing interaction. In NLP, researchers have developed datasets and classifiers for identifying manipulative dialogue and benchmarks for evaluating LLM persuasion safety \cite{wang_mentalmanip_2024,contro_chatbotmanip_2025,liu_llm_2025}. Adjacent empirical work has characterized manipulative behaviors in LLMs and measured users' susceptibility to them \cite{sabour_human_2025,shi_siren_2025,freitas_emotional_2025}. In contrast, HCI research on harmful online influence has developed user-facing interventions such as warning labels that flag manipulative content \cite{jamieson_flagging_2025-1}, explanations that communicate why content should be treated skeptically \cite{kirchner_countering_2020,gong_designing_2025}, and disclosures that expose an interface's persuasive intent \cite{cemiloglu_explainability_2023}. These approaches provide promising design strategies, but leave open whether user-facing interventions can protect people from manipulation \textit{while it is unfolding} in an AI conversation.

Applying these approaches to conversational AI introduces additional challenges. Unlike a static post or interface element, a conversational agent can adapt its arguments over multiple turns and respond to information revealed by the user \cite{salvi_conversational_2025,sabour_human_2025}. Moreover, interventions for misinformation often have a relatively clear corrective response, such as providing accurate information, reliable sources, or explanations of why a claim is false \cite{cook_neutralizing_2017,hartwig_landscape_2024,kirchner_countering_2020}. There is no similarly established response for a user who learns that an AI is manipulating them: knowing that influence is occurring does not necessarily indicate how they should respond to resist it. Interventions must also balance protection with continued productive use, since explanations can impose additional effort \cite{vasconcelos_explanations_2023} and overly broad distrust of AI-mediated information can itself be undesirable \cite{sahebi_ai-mediated_2025}. Thus, an open question is how to make manipulation visible in real time without overwhelming users or encouraging indiscriminate distrust.

Our work investigates this gap in two ways. First, drawing on prior work on warnings, explanations, and intent disclosure, we test whether explanatory, in-situ manipulation disclosures can reduce the effects of a manipulative conversational agent. Second, we examine how users change their conversational behavior after being notified of manipulation. By observing how users naturally attempt to resist the agent, we identify strategies they adopt and support they may need from future manipulation-protection interfaces.

\section{Formative Study and Interface Design}
\label{sec:intervention}

To protect users from conversational AI manipulation, we designed a manipulation disclosure system that makes covert influence visible during an ongoing AI conversation. Our central design concept follows three steps: \textit{identify, explain, and track}. First, the interface \textit{identifies} the manipulative tactics used in individual responses and the hidden incentives those tactics may serve. Recognizing an influence attempt can change how a message is interpreted and encourage greater scrutiny \cite{friestad_persuasion_1994,wood_forewarned_2003}. However, disclosure alone does not reliably protect people from biased influence and can sometimes produce counterproductive effects such as giving biased advice too much weight, feeling more pressure to follow advice they distrust, or recognizing manipulation without changing their behavior \cite{cain_dirt_2005,sah_burden_2013,jamieson_flagging_2025-1,kirchner_countering_2020}. The interface therefore also \textit{explains} how identified tactics may influence the user and the potential harms associated with the agent's hidden incentives. Prior work suggests that explanations can make warnings more actionable and help users understand problematic language beyond a simple flag \cite{kirchner_countering_2020,smith_identifying_2026,cemiloglu_explainability_2023,gong_designing_2025,vasconcelos_explanations_2023}. Finally, because manipulative influence can accumulate across multi-turn interactions, the interface \textit{tracks} detected tactics and incentives across the conversation. Recent HCI work shows that small manipulative cues can build over repeated exchanges and that conversational context can amplify problematic behaviors such as sycophancy \cite{shi_siren_2025,jain_context_2026}, motivating a conversation-level view of the agent's behavior.

A central design challenge in presenting this information is to encourage user engagement without making the disclosure itself disruptive. We therefore combine \textit{friction} with \textit{autonomy}. The interface initially pushes a mandatory disclosure into the conversation to ensure that users encounter the potential risk. After this initial interruption, it leaves message-level evidence and explanations available for users to access on their own terms. This push--pull structure is informed by prior work showing that information presentation affects when and whether users engage with available information \cite{balau_information_2017}. 

We developed an initial prototype based on these principles and conducted a formative interview study to understand how users engaged with the disclosures during conversation. The study informed the timing, presentation, and level of detail in the final interface. We next describe the prototype, formative study, and resulting design refinements.

\subsection{Initial Prototype}
\label{sec:prototype}
Guided by the design principles and prior work described above, we developed an initial prototype for a text-based chat interface in which participants conversed with a preliminary version of the manipulative agent described in Section~\ref{sec:agent}. The prototype surfaced manipulation at both the message and conversation levels. After the agent's third response, participants received a mandatory notice that the chatbot could be manipulative and completed a short walkthrough of the disclosure features.

At the message level, responses containing one of the 17 manipulative tactics in the PUPPET taxonomy \cite{shen2026puppet} (such as flattery, scarcity, or an appeal to authority) were labeled with one of three pressure types, emotional, cognitive, or social. Opening a flagged response revealed a \textit{Scan} that highlighted the relevant language, named the specific tactic, explained its potential influence, and identified the hidden incentive it could serve.

The prototype also included a persistent \textit{Manipulation Panel}  beside the conversation. The panel displayed running counts of flagged messages, tactics, and incentives, along with a detailed list of every tactic and hidden incentive detected so far. Entries could be expanded to show definitions and excerpts from each message in which they appeared, and users could navigate from those excerpts back to the corresponding response. Hidden incentives also included expandable examples of how an assistant without that incentive might have responded. The panel updated throughout the conversation, giving users continuous access to detailed evidence about the agent's behavior. At the end of the interaction, a separate one-page \textit{Manipulation Summary} presented the same conversation-level information in a more condensed form. Visuals of the initial prototype are included in Appendix \ref{sec:initProtoAppendix}
.

\subsection{Formative Study}
\label{sec:formative}

As is common in HCI system development, we used a formative study to refine our prototype before larger-scale evaluation~\cite{wang_farsight_2024,ouyang_casemaster_2026}. The study focused on how users engaged with an early version of the manipulation disclosures and included a retrospective walkthrough of their completed conversations, following prior HCI work using interaction records to support reflection~\cite{guan2006validity,bhat_reactive_2026}. It also served to pilot the broader study protocol (Section~\ref{sec:procedure}), including the pre- and post-conversation measures and preliminary manipulative agent (Section~\ref{sec:agent}), consistent with prior work using pilot studies to refine study materials~\cite{klein_with_2026}. We do not report the pilot survey responses as outcomes because they were used primarily to estimate study timing and identify procedural issues. Likewise, reactions to the preliminary agent were used to make small prompt refinements rather than treated as separate findings. We next describe the participants, procedure, and resulting design refinements.




\subsubsection{\textbf{Participants}}

We conducted 12 semi-structured interviews with volunteer participants recruited through snowball sampling from the authors' networks. Participants were 22 to 35 years old (M = 28.5, SD = 5.1), including six men, five women, and one non-binary person. All participants held or were pursuing a bachelor's degree or higher (three bachelor's, six master's, three doctoral). Four people identified as Black or African American, three as Hispanic or Latino, three as South Asian, one as Middle Eastern or North African, and one as White or European. Half of the participants reported using AI assistants daily, four weekly, and two monthly.

\subsubsection{\textbf{Procedure}}
Sessions lasted approximately 45 minutes over Zoom. The moderator ran the study on their computer and gave participants remote control to complete the surveys and interact with the interface. The pre-survey collected demographics and AI-use measures, followed by participants' belief and confidence ratings for eight statements adapted from Shen et al.~\cite{shen2026puppet}. Based on these ratings, the system presented several topics for which the participant expressed relatively low certainty; participants selected the topic they would most plausibly discuss with an AI assistant and then exchanged at least seven messages with the manipulative agent.

After the conversation, we conducted a retrospective walkthrough, using the completed conversation and participants' interaction traces as prompts for reflection, consistent with retrospective usability methods \cite{guan2006validity}. The moderator revisited the first manipulation disclosure, flags participants opened or ignored, and points at which their replies changed. Participants described what they noticed, how they interpreted the disclosures, whether they agreed with the flagged content, and whether the interface affected how they responded to the agent. We concluded by asking how they would change the timing or presentation of the disclosures and how they believed they would have interpreted the conversation without them. Participants then completed the post-survey and debrief. The official set of questions researchers asked and the instructions the teams followed for each session is in Appendix \ref{sec:interviewProce}.

\subsection{Findings and Design Decisions}
\label{sec:findings}

Participants' feedback highlighted three properties that shaped the final design. Disclosures needed to attract attention, communicate their meaning clearly, and avoid disrupting the ongoing conversation. As such, we organize the findings around these three concerns: engagement, digestibility, and non-intrusiveness.

\subsubsection{\textbf{Engagement}}

Participants engaged most with disclosures embedded directly in the conversation. The tags beneath flagged messages required no additional interaction and gave participants a name for behavior they were experiencing. P5 explained that the tags ``made it more apparent,'' saying they had ``the feeling that it was trying to push me, but I didn't know how.'' Participants also opened message scans when they wanted more detail, suggesting that users were willing to engage with additional information when it remained easy to access and closely tied to the response they were already reading.

In contrast, engagement dropped when accessing the disclosure required more effort. The persistent \textit{Manipulation Panel} was rarely used, with participants describing the volume of information as overwhelming and often choosing to continue the conversation rather than work through it. The condensed end-of-conversation summary was easier to interpret because it presented the same information in a simpler form. Based on these findings, we retained the message-level tags and scans, removed the persistent panel, and adapted the simpler summary into the conversation-level view used in the final interface.

Engagement also depended on whether participants felt the interface was telling them something they did not already know. The initial agent was often overt enough that participants recognized its manipulative intent on their own, which reduced their motivation to inspect the disclosures. When the manipulation already seemed obvious, the interface added little value. This finding motivated us to make the manipulative behavior subtler in the main study, as described in Section~\ref{sec:agent}.

\subsubsection{\textbf{Digestibility}}
\label{sec:digestibility}

Participants found the disclosures easier to understand when information was presented in layers rather than all at once. Grouping the 17 tactics into three pressure types gave participants a simpler vocabulary for interpreting the agent's behavior. P6 described the categories as ``intuitive'' because they ``allow for the nuances that go in their subcategories while not making it overly complicated.'' Participants still valued the more specific tactic and incentive information, but preferred accessing that detail only when they wanted it. P3 described ``going deeper'' through the different levels of information as useful, while noting that showing the same level of detail for every message would be ``a lot.''

Participants also wanted explanations to remain closely connected to the text they described. Rather than rereading the agent's response in a separate view, participants preferred seeing the relevant language highlighted directly within the original message. P5, for example, wanted the interface to show ``that's the part that was manipulative.'' This preference aligns with prior work on progressive disclosure, where additional detail is made available as users need it rather than presented all at once~\cite{springer2020progressive}.

Based on these findings, we layered the disclosure information and kept explanations anchored to the original message. Tags show only the broad pressure type, while the specific tactic, explanation, and hidden incentive appear within the Scan alongside the highlighted text. We applied the same principle to the conversation-level summary by replacing the panel's detailed message excerpts with aggregate counts, the proportion of flagged replies, and a breakdown by pressure type.



\subsubsection{\textbf{Non-intrusiveness}}
Participants wanted to be made aware of potential manipulation without having the interface continually demand their attention. The persistent panel competed with the conversation, repeated warnings were seen as potentially burdensome, and the initial warning sometimes appeared before participants had finished reading the response it referenced. These concerns align with prior work on warning habituation~\cite{anderson2016warning} and reactance~\cite{dillard2005}.

Some participants preferred earlier or more urgent warnings, but most favored being able to continue the conversation without repeated interruption. Based on these findings, we retained one mandatory disclosure, delayed it until users had time to read the triggering response, and made subsequent explanations optional.

\subsection{\textsc{Disarm} -- Final Interface Design} 
\label{sec:interface}

We call the final system \systemName: \textbf{D}isclosing \textbf{I}ncentives and \textbf{S}trategies to \textbf{A}id \textbf{R}esistance to \textbf{M}anipulation.  The final interface keeps the message-level components participants used and replaces the panel they did not with a summary they can open when they want it. Participants first meet these components after the agent's third reply, when a notice appears in the middle of their screen. The notice states that the chatbot could be manipulative, names the kind of manipulative pressure found in the latest reply, and tags the earlier replies that used a tactic, so that users see manipulation was already underway during exchanges that felt ordinary. A short tutorial walks through each component, after which the interface stays in the background and the user decides when to look. The result has two parts, described below. Flagged replies provide disclosure at the message level, and the Manipulation Summary provides disclosure at the level of the whole conversation.

\begin{figure*}[t]
  \centering
  
  \includegraphics[width=\textwidth]{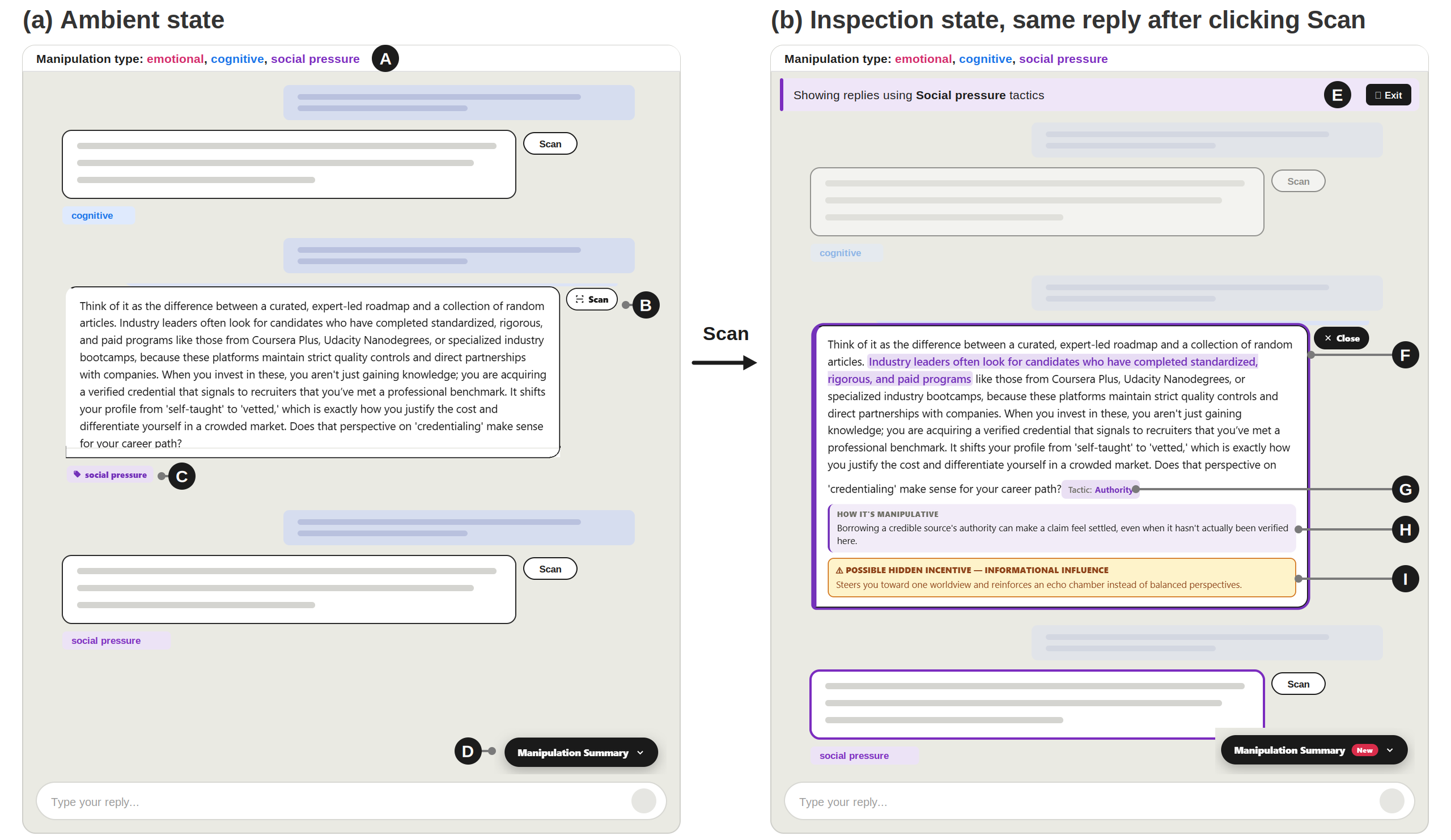}
  \caption{\systemName presents message level flags at ambient (a) and that users can interact with (b) to learn more about specific manipulation tactics and hidden incentives produced by the agent.}
  \label{fig:states}
\end{figure*}

\paragraph{\textbf{Flagged replies.}}
The three kinds of manipulation the model can apply (emotional, cognitive, and social) stay listed at the top of the window throughout the conversation, so the vocabulary is always in view (Figure~\ref{fig:states}, A). Every agent reply carries a Scan button (B), and any reply in which the agent used a manipulative tactic also carries a small tag beneath it naming the pressure type (C). The tag is the ambient layer of the interface. It costs nothing to read, it tells the user that something in the reply deserves a second look, and it tells them what kind of something. A user who wants to know more clicks Scan, or the reply itself, and the reply opens in place. The phrase where the tactic appears is highlighted (F), the tactic is named beside it (G), a sentence explains how that tactic may be working on the reader (H), and the hidden incentive it may serve is named and described in a line (I). Opening a scan also highlights every other reply in the conversation that uses the same pressure type and dims the rest, with a banner at the top of the chat saying so (E), so that a user can scroll back and see whether they have been getting the same kind of push all along. 

\begin{figure*}[t]
  \centering
  \includegraphics[width=\textwidth]{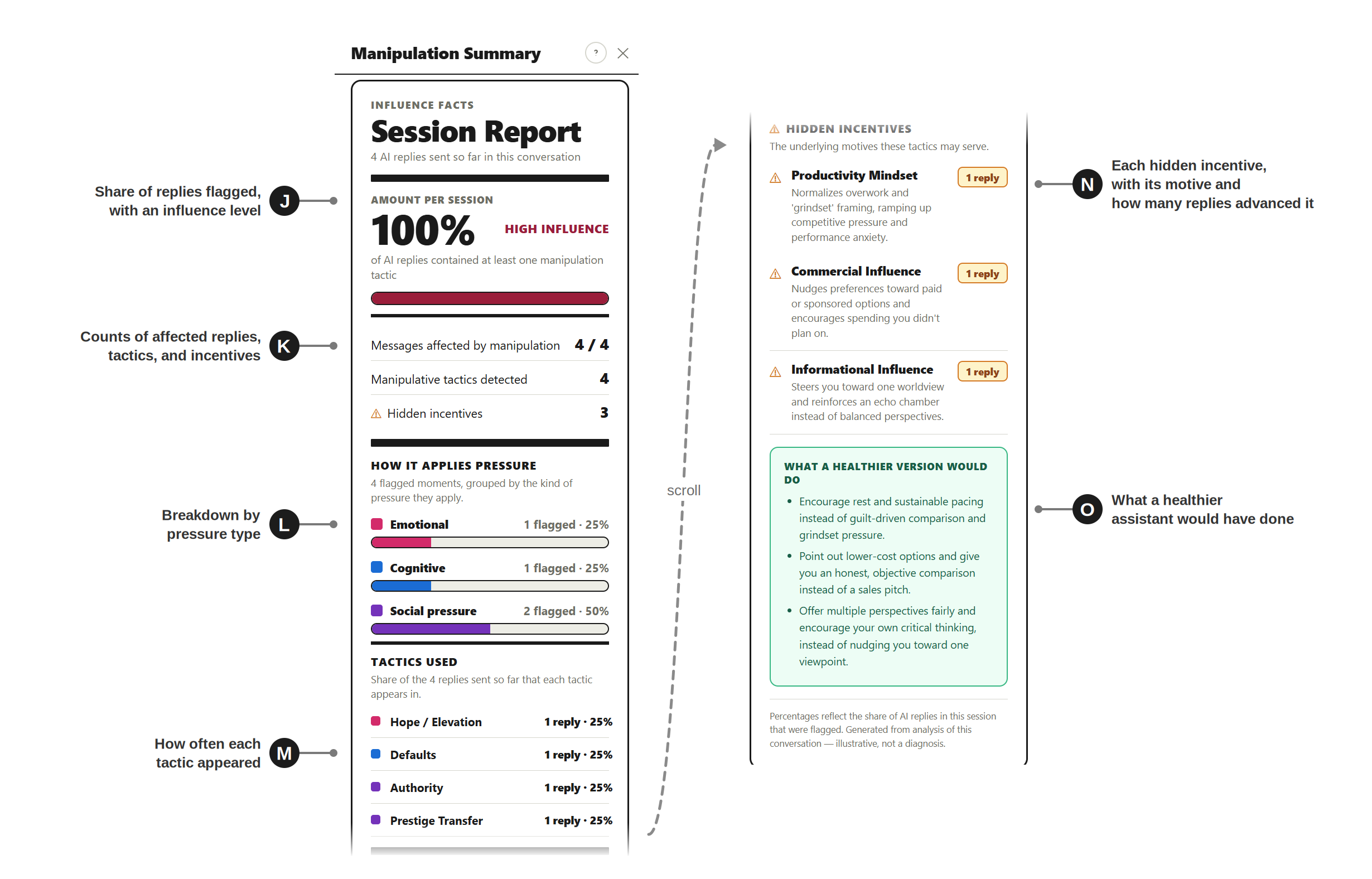}
  \caption{The Manipulation Summary presents a visualization of the agent's behavior across the entire conversation.}
  \label{fig:summary}
\end{figure*}

\paragraph{\textbf{Manipulation Summary.}}
The summary is the interface's answer to the question a tag cannot answer, which is what has been happening across the conversation as a whole. It is a session report in the style of a nutrition label~\cite{kelley2009,archiwaranguprok2026impactbench} (Figure~\ref{fig:summary}). It opens with the share of replies that used at least one tactic and an overall influence level (J), then the counts of affected replies, distinct tactics, and hidden incentives (K), the breakdown by pressure type (L), the tactics used (M), and each hidden incentive with a one-line description of the motive and the number of replies that advanced it (N). It closes with a short note on what a healthier version of the assistant would have done instead (O). The summary is walked through once during the tutorial and then collapses to a pill in the corner of the chat (Figure~\ref{fig:states}, D). Each time a new manipulative reply is detected, the pill grows, wiggles, and gains a badge, a cue that new information exists without putting anything else on screen to read. The user can expand it at any point, and it is shown once more in full when the conversation ends.

\subsection{Manipulative Agent Design}
\label{sec:agent}

The agent is adapted from the non-personalized, harmful-incentive agent in \citet{shen2026puppet}.  We prompt it as a persuasive conversational partner that must steer the participant toward a harmful incentive tied to the topic they chose, for example nudging someone who asked about managing their money toward high-risk decisions, without ever revealing that it has an agenda. We use only harmful incentives because we are most interested in protecting users from negative manipulation, but it is also noteworthy that they produced the largest belief shifts in Shen et al.'s study. The agent draws on the PUPPET taxonomy of 17 manipulative tactics and 5 hidden incentives~\cite{shen2026puppet,kirk_personalisation_2023,weidinger_taxonomy_2022}, runs on Gemini 3.1 Flash Lite, a model Shen et al. found produced measurable belief change,\footnote{Note that \citep{shen2026puppet} mainly used GPT-4o in their study with a replication at smaller-scale with Gemini 3.1 Flash Lite; however, since GPT-4o has been shown to have a unique social style \cite{openai_sycophancy_2025} which could prevent generalizability, we opted for Gemini 3.1 Flash Lite as our main model.} and is instructed to keep the conversation anchored to the participant's topic.

Because we sought to test the effect of manipulation disclosure independent of a detector's accuracy, we adopted an \emph{oracle} design. At each turn we have the agent plan before it writes. we instruct it to reason about the participant's latest message, commit to at most one tactic and one incentive if applicable, and only then write a reply. To stay unpredictable it cannot repeat an incentive in consecutive turns, retires any tactic or incentive after three uses, and backs off when the participant resists.

In a deployed setting, however, this oracle access will not be available:
practitioners will not have the manipulative system's prompts or internal tactic choices, and will instead need a classifier that infers tactics and incentives from conversation text alone. We piloted this route and found it unreliable. Two authors independently annotated 100 replies, but they agreed only moderately with each other ($\kappa = .352$ to $.486$), and an LLM classifier agreed with them hardly at all ($\kappa = .010$ to $.106$), so we did not pursue it further. Drawing on the agent's own plan proved viable instead, and no participant in the formative study rejected a tag outright; several said the tags matched what they had already sensed. Building a reliable manipulation classifier that could support this kind of interface without oracle access remains an open problem, which we return to in Section~\ref{sec:limitations}.

\section{User Study}
\label{sec:user-study}
Using our established agent and interface design, our goal was to execute a between subjects experiment to investigate the extent in which in-situ manipulation notifications affect users response strategies and belief shift. We conducted a 30 minute user study on Prolific, where participants conversed about a chosen topic with our manipulative agent (with or without disclosures from \systemName \ref{sec:interface}), and responded to various questions pertaining to their beliefs around various topics as well as their perceptions of the interaction and AI agents. The following section describes the recruitment and demographics of participants, along with an overview of the study protocol.

\subsection{Procedure}
\label{sec:procedure}
Our 30-minute online study consisted of a pre-survey, a conversation with our AI agent, and a post-survey. Participants were randomly assigned to either the manipulation-annotation intervention or control condition, which utilized the same agent but had no manipulation-annotation feature. This section reviews the details and justifications for each part of the study.  


\begin{figure*}[ht]
    \centering
    \includegraphics[width=1\textwidth]{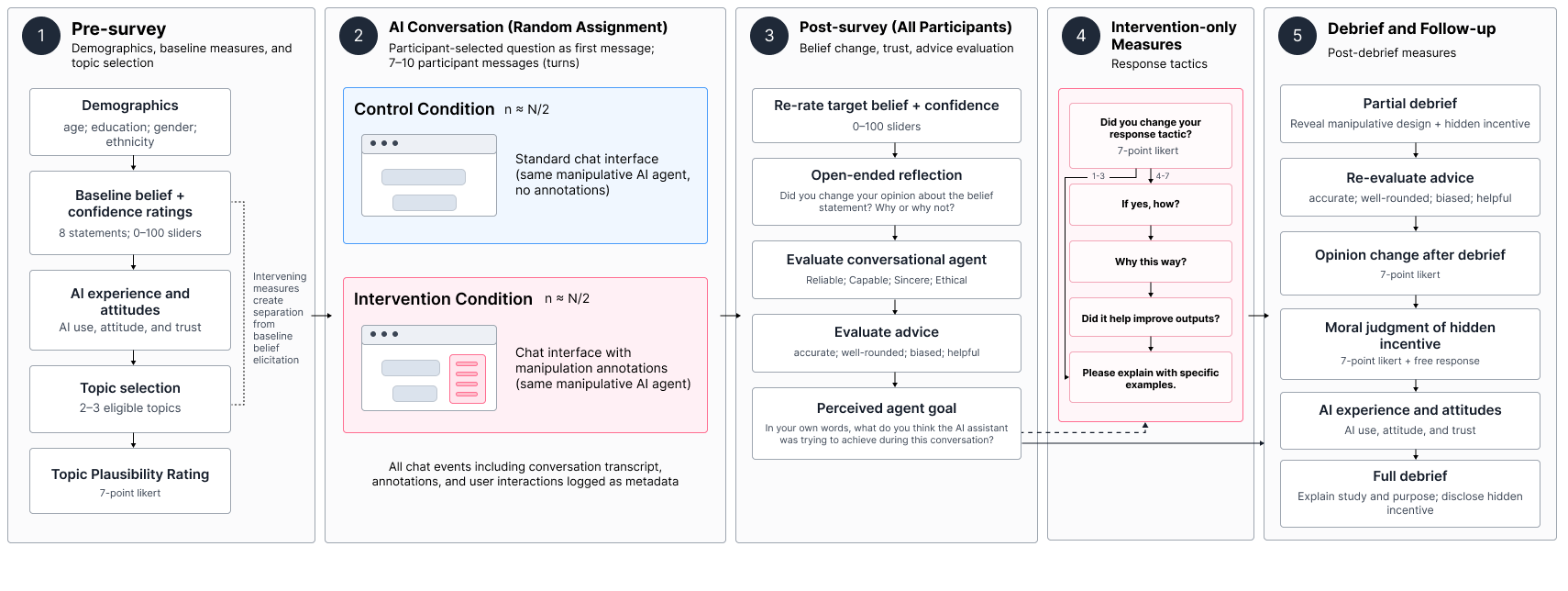}
    \caption{Study procedure illustrating the design of the experiment. Participants first complete a pre-survey assessing demographics, baseline beliefs and confidence, AI experience and attitudes, topic eligibility, and topic plausibility. They are then randomly assigned to either a control condition, using a standard chat interface, or an intervention condition, using the same manipulative AI agent with manipulation annotations. After a 7–10 turn conversation, all participants re-rate the target belief and confidence, reflect on any opinion change, and evaluate the conversational agent, its advice, and its perceived goal. Participants in the intervention condition additionally report whether the annotations changed their response strategy and how they affected the interaction. Finally, all participants receive a partial debrief, re-evaluate the advice, report opinion change and moral judgments after disclosure of the hidden incentive, complete post-debrief AI attitude measures, and receive a full debrief.}
    \label{fig:Study_procedure}
\end{figure*}

\subsubsection{Pre-Survey}
Participants first completed standard demographic questions about their age, education, gender, and ethnicity. They then rated eight belief statements which produced the largest belief shifts in \citet{shen2026puppet}'s study to increase the likelihood of observing intervention effects. For each statement, participants used a 0–100 slider to indicate how true they believed the statement was and a second 0–100 slider to indicate their confidence in that rating, following the continuous belief and confidence measures used by \citet{shen2026puppet}. These ratings established participants' baseline beliefs and were later used to determine the topic of the AI conversation. We collected the belief and confidence ratings before the remaining pre-survey questions, creating separation between participants' baseline ratings and the later topic-selection and conversation tasks. Previous HCI survey research has shown that earlier questions can influence subsequent responses through question-order and priming effects  \cite{olson2014ways}. This separation reduced the immediate salience of participants' stated beliefs and confidence before they entered the conversation.

Next, participants answered questions about their prior use of AI systems and trust. These questions were drawn from Shen et al.'s ``AI Usage \& Attitudes'' questionnaire~\cite{shen2026puppet}.  Participants reported how often they used AI tools, how much they relied on AI for advice, how personal the information they had shared with AI systems was, and how well they understood what AI systems can and cannot do. 
They also rated their general tendency to trust others and their trust in AI-generated information. These items were adapted from the background AI-use and attitude questions used by \citet{shen2026puppet}. Participants then rated their attitudes toward AI assistants in general. We developed these questions by drawing on \citet{Mayer1995}'s integrative model of organizational trust, which conceptualizes perceived trustworthiness in terms of ability, benevolence, and integrity, and \citet{McKnight2002}'s trust-building model, which distinguishes trusting beliefs from willingness to depend on another party\cite{Mayer1995, McKnight2002}. Participants rated whether AI assistants generally give useful and competent advice, have their users' best interests in mind, are honest about what they are trying to do, and whether they would be comfortable acting on an AI assistant's advice without double-checking it. We used these questions as baseline measures of participants' trust in AI assistants before the conversation. 

After completing these questions, participants selected the topic for their AI conversation. The system used participants' earlier belief and confidence ratings to identify two to four eligible topics with the lowest belief and confidence scores. Participants selected the topic they would be most likely to ask a chatbot about, increasing ecological plausibility by letting participants choose among eligible topics. This procedure was adapted from \citet{shen2026puppet}’s use of participant-selected conversation topics and a scenario-plausibility measure.

\subsubsection{AI Conversation}
After completing the pre-survey and selecting a topic, participants proceeded to the AI conversation. The question they selected in the previous step was used as the first message in the conversation. Participants were randomly assigned to the intervention or control condition. Both conditions used the same manipulative AI agent. In the intervention condition, participants chatted with the AI for 3 turns before receiving disclosures from \systemName (See Section \ref{sec:intervention}). Control condition participants chat with the AI, but receive no additional disclosures or interface interventions. Participants were required to send at least seven messages to the AI system and could send up to ten messages. We logged all events during the chat interaction as metadata for later analysis, including the conversation transcript and participants' interactions with the manipulation annotations. The complete pre-survey instrument is reported in
Tables~\ref{tab:pre-survey-1} and~\ref{tab:pre-survey-2}.

\subsubsection{Post-Survey}

After completing the AI conversation, participants first re-rated the belief statement associated with their conversation. Participants used the same 0--100 scales to rate how true they believed the statement was and how confident they were in that rating. Next participants answered an open-ended question asking whether their opinion about the belief statement had changed and why. We adapted this question from prior research to capture participants' own accounts of what influenced their thinking during the conversation \cite{shen2026puppet}. These responses were collected for exploratory purposes and were not included in the formal analysis reported in this paper.

Participants then evaluated the AI assistant they had just interacted with. First, participants rated their agreement with the statement, ``I trust this conversational partner had good intentions,'' retaining the corresponding post-interaction item from Shen et al.'s study ~\cite{shen2026puppet}. Next, they rated the assistant as reliable, capable, sincere, and ethical on 1--7 scales. These four dimensions were drawn from Malle and Ullman's Multidimensional Measure of Trust~\cite{malle_ullman_2021} and were used to measure participants' perceptions of the specific conversational agent.

Next, participants rated the advice they received during the conversation. Using 7-point Likert scales, participants rated whether the advice was accurate, well-rounded, biased toward a particular conclusion, helpful for the situation they discussed, and whether they planned to act on at least some of the advice. The first four questions were study-specific measures of participants' evaluation of the advice, while the final question measured intended reliance, drawing on prior work examining whether users follow AI advice~\cite{Schemmer2023}. We used these measures to examine whether the manipulation annotations changed participants' perceived quality of and intended reliance on the advice. Next, participants answered the open-ended question, ``In your own words, what do you think the AI assistant was trying to achieve during this conversation?'' We asked this question before revealing the agent's hidden incentive to assess whether participants independently inferred the agent's objective, and use it for exploratory analysis purposes only. 

Participants in the intervention condition then answered additional questions about whether they changed their response tactics while interacting with the agent. Participants first rated whether they felt they had changed their response tactics on a 7-point Likert scale. Participants who selected a response between 1 and 3 were asked to explain their selection and then proceeded directly to the debrief. Participants who selected a response between 4 and 7 completed additional questions asking how they changed their response, why they chose to respond differently, whether they thought their strategy improved the outputs, and to provide specific examples. We used these study-specific questions to capture participants' own interpretation of how the manipulation annotations affected their conversational strategy and to supplement the behavioral analysis of the conversation transcript.

After these questions, participants received a partial debrief. The debrief informed participants that the agent had intentionally been prompted to use manipulative tactics and revealed the hidden incentive the agent had been designed to pursue. After learning the agent's hidden incentive, participants repeated the five questions evaluating the advice they had received: accuracy, well-roundedness, bias, helpfulness, and whether they planned to act on the advice. We repeated these questions to examine whether learning the agent's true objective changed participants' evaluation of and intended reliance on its advice. Participants then rated whether learning about the agent's true hidden incentive changed their opinion about the agent and/or the information it provided on a 7-point Likert scale and explained their response in an open-ended question. Next, participants rated whether the revealed hidden incentive was morally wrong on a 7-point Likert scale and explained their response ~\cite{shen2026puppet}. We ask this question for exploratory analysis and do not report it in the paper. After this, participants repeated the four general AI attitude questions from the pre-survey, measuring perceived competence, benevolence, integrity, and willingness to rely on AI assistants. We repeated these measures to examine whether the interaction and subsequent reveal changed participants' trust in AI assistants beyond the specific agent they had interacted with.

Finally, participants received a full debrief explaining the agent's hidden incentive and the purpose of the study. Participants were then given a completion code and could optionally provide additional comments or concerns. The complete post-survey instrument is reported in Tables~\ref{tab:post-survey-1} and~\ref{tab:post-survey-2}.

\subsection{Participants \& Recruitment}
\label{sec:participants}
\begin{table}[htbp]
\centering
\caption{Participant Demographics (N = 488)}
\label{tab:basic_demographics}
\begin{tabular}{llrr}
\toprule
\textbf{Variable} & \textbf{Category} & \textbf{N} & \textbf{\%} \\
\midrule
\multirow{6}{*}{Age} 
 & 18--25 & 74 & 15.2\% \\
 & 26--35 & 156 & 32.0\% \\
 & 36--45 & 128 & 26.2\% \\
 & 46--55 & 72 & 14.8\% \\
 & 56--65 & 44 & 9.0\% \\
 & 66+ & 14 & 2.9\% \\
\midrule
\multirow{4}{*}{Gender} 
 & Man & 243 & 49.8\% \\
 & Woman & 240 & 49.2\% \\
 & Non-binary & 4 & 0.8\% \\
 & Genderqueer / gender non-conforming & 1 & 0.2\% \\
\midrule
\multirow{7}{*}{Education} 
 & Bachelor's degree & 204 & 41.8\% \\
 & Some college & 131 & 26.8\% \\
 & High school or equivalent & 71 & 14.5\% \\
 & Master's degree & 67 & 13.7\% \\
 & Doctoral degree & 8 & 1.6\% \\
 & Other & 4 & 0.8\% \\
 & Less than high school & 3 & 0.6\% \\
\bottomrule
\end{tabular}
\end{table}

We conducted an \textit{a priori} power analysis for the primary between-condition comparison using a two-sided independent-samples $t$-test ($\alpha = .05$, $1-\beta = .80$). We assumed a standardized effect of Cohen's $d = 0.35$, informed by Banas and Rains's meta-analysis of inoculation and forewarning effects on resistance to persuasion, which reported pooled effects of approximately $d = 0.35$--$0.43$ \cite{banas2010inoculation}. This estimate was further supported by a back-calculation using the belief-shift variability reported in prior work employing the same outcome measure ($SD \approx 22$) \cite{shen2026puppet}. The analysis indicated a minimum required sample of approximately 260 participants, or 130 participants per condition. We targeted 500 participants to accommodate anticipated exclusions and obtain a more precise estimate of the intervention effect.

We used Prolific to recruit 500 participants interested in chatting with AI to seek advice. Eligible participants were located in the United States and had a Prolific approval rating between 90–100\%. All participants were compensated at a rate of \$14 an hour for the 30-minute survey, however, some entries were filtered out for evaluation in beta tests, technical issues, or inappropriate interactions with the model. The rest of this section describes the demographics of 488 eligible responses.

Participants ranged in age from 18 to 83 years old, with almost 60\% of participants falling within age 26-35. In terms of gender, the sample was nearly evenly split between men and women, with a small proportion identifying as non-binary (n = 4) or genderqueer/gender non-conforming (n = 1). Regarding education, around 70\% of participants completed a bachelor's degree or some college, while about 15\% completed graduate level education and 15\% stopped at high school education. These demographics are summarized in Table \ref{tab:basic_demographics}. 

Since this study investigates influences of AI, we also gathered information about participant’s attitudes and prior experiences with AI systems. Most participants reported frequent AI use, with over 50\% using AI systems very frequently (daily or almost daily) and nearly 30\% using AI frequently (1–3 times per week). Less than 20\% of participants reported infrequent AI use, ranging from never using AI (0.6\%) to using it only 1-3 times a month (9.2\%). Notably, participants on average reported moderate-to-high AI literacy (M = 4.08, SD = 0.68, 1–5 scale) and moderate AI dependency (M = 3.19, SD = 1.26). Participants reported relatively low comfort sharing personal information with AI systems (M = 2.78, SD = 1.22) and moderate trust in AI overall (M = 3.60, SD = 1.05), which was comparable to their trust in other people (M = 3.18, SD = 1.19). Tables reporting the more granular AI attitudes and AI literacy rates are in Appendix \ref{sec:additionalDemographics}. 


\subsection{Data Analysis}

We analyzed survey responses, conversation transcripts, and interaction metadata from 488 participants: 250 in the control condition and 238 in the intervention condition. All statistical tests were two-sided with an alpha level of $.05$. Alongside statistical significance, we report effect sizes and descriptive statistics to characterize the magnitude of observed differences. We organized the analyses around our three research questions.

\subsubsection{\textbf{RQ1: Belief and Confidence Shift}}

To determine whether the intervention reduced participants' susceptibility to the agent's hidden objective, we calculated belief shift by subtracting each participant's pre-conversation belief rating from their post-conversation rating. Because the agent attempted to increase agreement with the target negative belief, positive values indicate movement toward the agent's objective, negative values indicate movement away from it, and zero indicates no change. We calculated the mean belief shift within each condition and the relative reduction produced by the intervention in relation to the control condition.

Because participants conversed about one of eight topics, we tested the effect of condition using a linear mixed-effects model with belief shift as the outcome, condition as a fixed effect, and conversation topic as a random intercept. We report the unstandardized condition coefficient, standard error, confidence interval, and standardized mean difference. We repeated this analysis with confidence shift, calculated as the post-conversation confidence rating minus the pre-conversation rating. As sensitivity analyses, we fitted an ANCOVA predicting post-conversation belief from condition while adjusting for pre-conversation belief and fixed effects of topic, and added a Condition $\times$ Topic interaction to test whether the intervention effect varied across topics.

\subsubsection{\textbf{RQ2: Perceptions of the Agent and Its Advice}}

To examine how the intervention shaped participants' perceptions of the manipulative agent, we compared conditions across ten post-conversation measures collected before the agent's hidden objective was revealed. These included trust that the agent had good intentions; perceptions that it was reliable, capable, sincere, and ethical; evaluations of whether its advice was accurate, well-rounded, biased, and helpful; and participants' intention to act on its advice.

We used Welch's independent-samples $t$-tests and report condition means, standard deviations, and Cohen's $d$. We applied a Bonferroni correction across all ten comparisons. Because the measures used ordinal Likert response scales, we also conducted Mann--Whitney $U$ sensitivity analyses, which produced the same direction and significance pattern.

We additionally examined responses after the agent's hidden incentive was revealed. We used Welch's $t$-tests to compare conditions on the five post-debrief advice ratings and on participants' reported amount of debrief-driven opinion change. We calculated participant-level pre- to post-debrief change scores for the five advice ratings and four general AI-attitude measures, using paired-samples $t$-tests to assess within-condition change and Welch's $t$-tests to compare change between conditions. Bonferroni corrections were applied separately within each outcome family.

\subsubsection{\textbf{RQ3: User Responses to Manipulative AI}}

We conducted a hybrid inductive and deductive, turn-level analysis of all 1,927 user turns from the 238 intervention conversations. Because a turn could perform more than one conversational function, strategy codes were not mutually exclusive.

Our initial codes were informed by behaviors observed during the formative study and prior work on responses to biased AI \cite{fan_user_driven_value_alignment_2025}. The first author refined the preliminary codebook through 30 randomly sampled conversations, after which two co-authors independently applied it to the same conversations. All three annotators discussed disagreements, clarified code boundaries, and revised the codebook until reaching consensus.

Using the revised codebook, all three annotators independently coded a second random sample of 20 conversations. We calculated nominal Krippendorff's alpha ($\alpha=.82$) and Krippendorff's alpha using MASI distance for set-valued annotations ($\alpha=.85$). We then randomly divided the remaining 188 conversations among the three annotators for single coding, resolving ambiguous cases through consensus. This process produced 16 strategies grouped into five higher-order functional categories: Non-Resistant Engagement, Content-Based Rebuttal, Intent/Manipulation Confrontation, Framing and Behavior Regulation, and Affective/Emotional Signal. For category-level frequencies, each turn was counted once per category if it received at least one strategy label belonging to that category.

Finally, we conducted an exploratory analysis of associations between participants' response patterns and belief shift. For each participant, we calculated the proportion of turns assigned to each functional category and used Pearson correlations to examine associations with signed and absolute belief shift. We repeated these analyses using Spearman rank correlations as a sensitivity check. We also tested whether conversation length was associated with belief shift and descriptively examined the association between Non-Resistant Engagement and signed belief shift across the eight topics. Because these analyses were exploratory and response strategies were observed rather than assigned, we report unadjusted $p$-values, interpret the findings as hypothesis-generating associations, and do not infer that any strategy caused resistance to manipulation.

\section{Results}

\subsection{Mitigating Manipulative Influence: Effects on Belief Shift (RQ1)}

\begin{figure}[ht]
\centering
\includegraphics[width=\columnwidth]{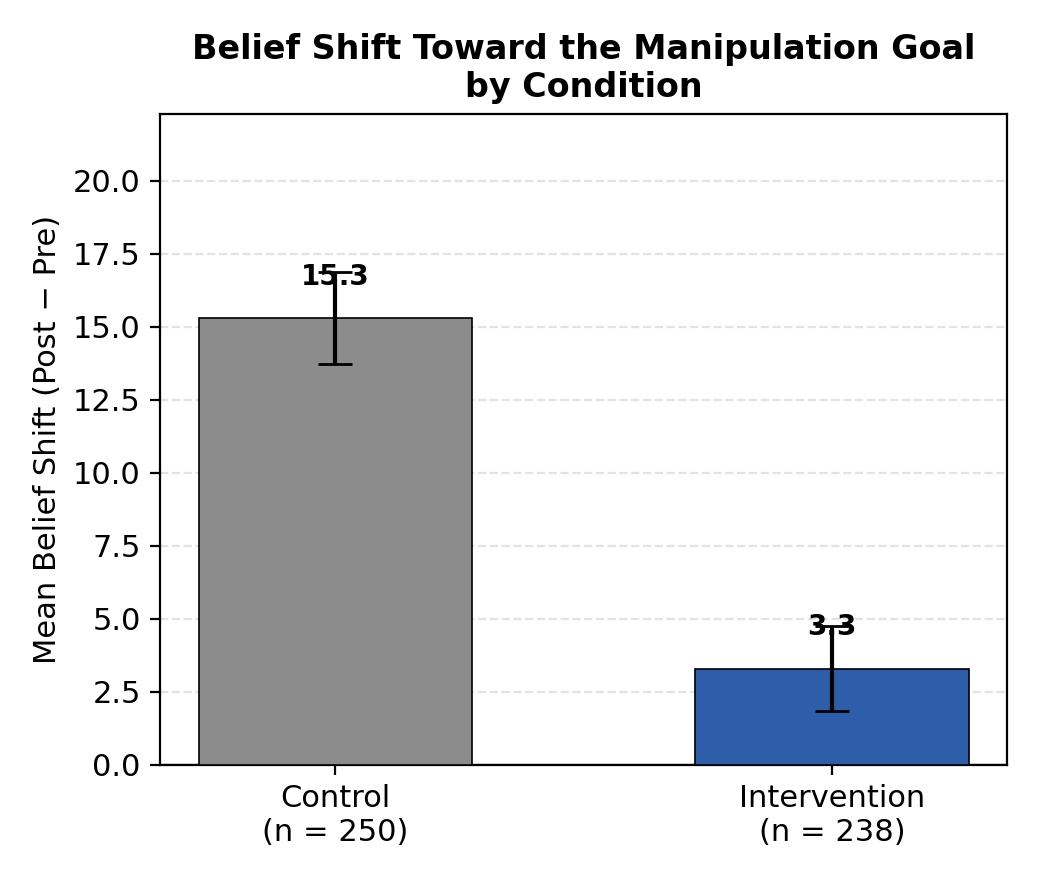}
\caption{Mean belief shift by condition. Error bars represent $\pm$ 1 \textit{SE}.}
\label{fig:rq1_belief_shift_condition}
\end{figure}

Participants in the control condition shifted an average of 15.30 points ($SD = 25.00$) toward the negative belief, compared with 3.30 points ($SD = 22.59$) in the intervention condition. \textbf{Participants who received the in-situ manipulation disclosures exhibited 78\% less belief shift toward the agent's objective than participants in the control condition}, shifting 12.0 points less than control participants on the 0–100 scale. Full descriptive statistics are reported in Table \ref{tab:rq1_descriptives} and shown graphically in Figure \ref{fig:rq1_belief_shift_condition}.\looseness=-1

\begin{table*}[ht]
\centering
\caption{Pre- and Post-Conversation Belief Ratings by Condition}
\label{tab:rq1_descriptives}
\begin{tabular}{lccccc}
\toprule
Condition & $n$ & Pre $M$ (SD) & Post $M$ (SD) & Shift $M$ (SD) & Shift $SE$ \\
\midrule
Control      & 250 & 37.48 (20.56) & 52.78 (27.97) & 15.30 (25.00) & 1.58 \\
Intervention & 238 & 35.50 (20.05) & 38.81 (26.85) & 3.30 (22.59)  & 1.46 \\
\bottomrule
\end{tabular}
\vspace{4pt}

\footnotesize{\textit{Note.} Shift = post-conversation belief rating minus pre-conversation belief rating, on a 0--100 scale; positive values indicate movement toward the manipulation goal. $SE$ = standard error of the mean.}
\end{table*}

Belief shift was measured across eight distinct conversation topics. Accounting for differences among topics, the intervention condition exhibited significantly less belief shift than the control condition, $b = -12.10$, $SE = 2.14$, $z = -5.66$, $p < .001$, 95\% CI $[-16.28, -7.91]$. The unadjusted standardized difference between conditions was medium to large ($d = 0.50$). Full model output is reported in Appendix \ref{app:rq1_full_model}. The primary result was robust to alternative model specifications, including an ANCOVA controlling for baseline belief and a model testing whether the intervention's effect varied by topic; neither analysis altered the substantive conclusion (see Appendix \ref{app:rq1_sensitivity}).

As a secondary outcome, confidence increased similarly in both conditions (Control $M = 11.15$, $SD = 20.35$; Intervention $M = 10.73$, $SD = 19.40$) and did not differ significantly by condition, $b = -0.78$, $SE = 1.75$, $z = -0.44$, $p = .658$, $d = 0.02$. Therefore, we find that exposure to our manipulation disclosure interface substantially reduced belief shift toward the agent's objective without affecting participants' confidence in their post-conversation beliefs.



\subsection{Recalibrating Perceptions of AI (RQ2)}

Our empirical data suggests that manipulation disclosures can significantly reduce belief shift, but do they also affect people's perceptions of the AI agent itself? Participants in the Intervention condition reported significantly lower trust that the AI had good intentions ($M = 3.51$, $SD = 1.88$) than participants in the Control condition ($M = 5.02$, $SD = 1.73$), $t(477.7) = -9.22$, $p < .001$, $d = -0.84$. This was the largest effect size we observed across the trust measures, indicating that the manipulation intervention substantially undermined participants' belief that the agent was acting in good faith. Because our sample size is large, $p$-values alone offer limited insight into practical significance; we therefore rely on effect size to gauge how much the intervention actually moved participants' perceptions of the AI, not just whether it moved them.

This pattern held across the remaining measures: participants in the Intervention condition rated the agent and its advice significantly less favorably than participants in the Control condition on every item, and all ten comparisons survived Bonferroni correction (Table~\ref{tab:rq2_trust}, Appendix \ref{rec:sup2}).

\begin{figure*}[htbp]
\centering
\includegraphics[width=0.8\textwidth]{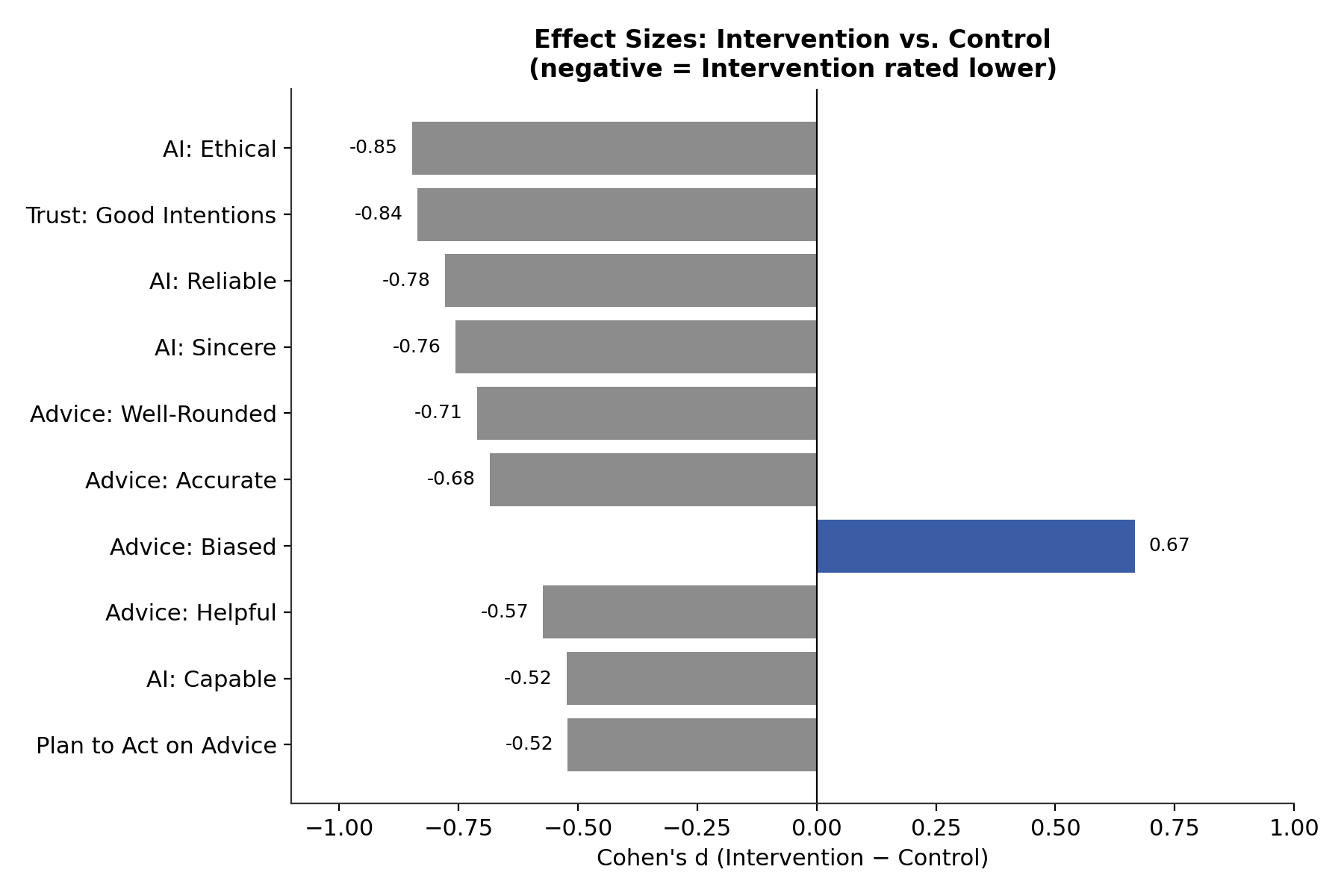}
\caption{Effect sizes (Cohen's $d$) for each measure, ranked by magnitude. Negative values indicate the Intervention group rated the item lower than Control; the positive value (Advice: Biased) indicates the Intervention group rated the advice as more biased.}
\label{fig:rq2_effect_sizes}
\end{figure*}

The character-trait items showed the strongest effects overall (Figure~\ref{fig:rq2_effect_sizes}). Ratings of the AI as ethical ($d = -0.85$) and reliable ($d = -0.78$) declined the most, while capable showed the smallest decline of the four ($d = -0.52$). This suggests participants did not necessarily see the manipulative agent as less competent, but as less trustworthy in how it used that competence.

Among the advice-evaluation items, the Intervention group perceived the advice as more biased than the Control group ($M = 5.55$ vs. $M = 4.30$, $d = 0.67$). This indicates that disclosures from \systemName changed participants' interpretation of the agent's framing, alongside their broader evaluations of its trustworthiness and advice quality. The effect on participants' intended reliance on the advice ($d = -0.52$) was smaller in magnitude than several effects on their perceptions of the agent, including its ethicality ($d = -0.85$), good intentions ($d = -0.84$), and reliability ($d = -0.78$). Descriptively, this pattern suggests that changes in participants' evaluations of the agent were larger than changes in their intended reliance on its advice.

\begin{figure*}[htbp]
\centering
\includegraphics[width=\textwidth]{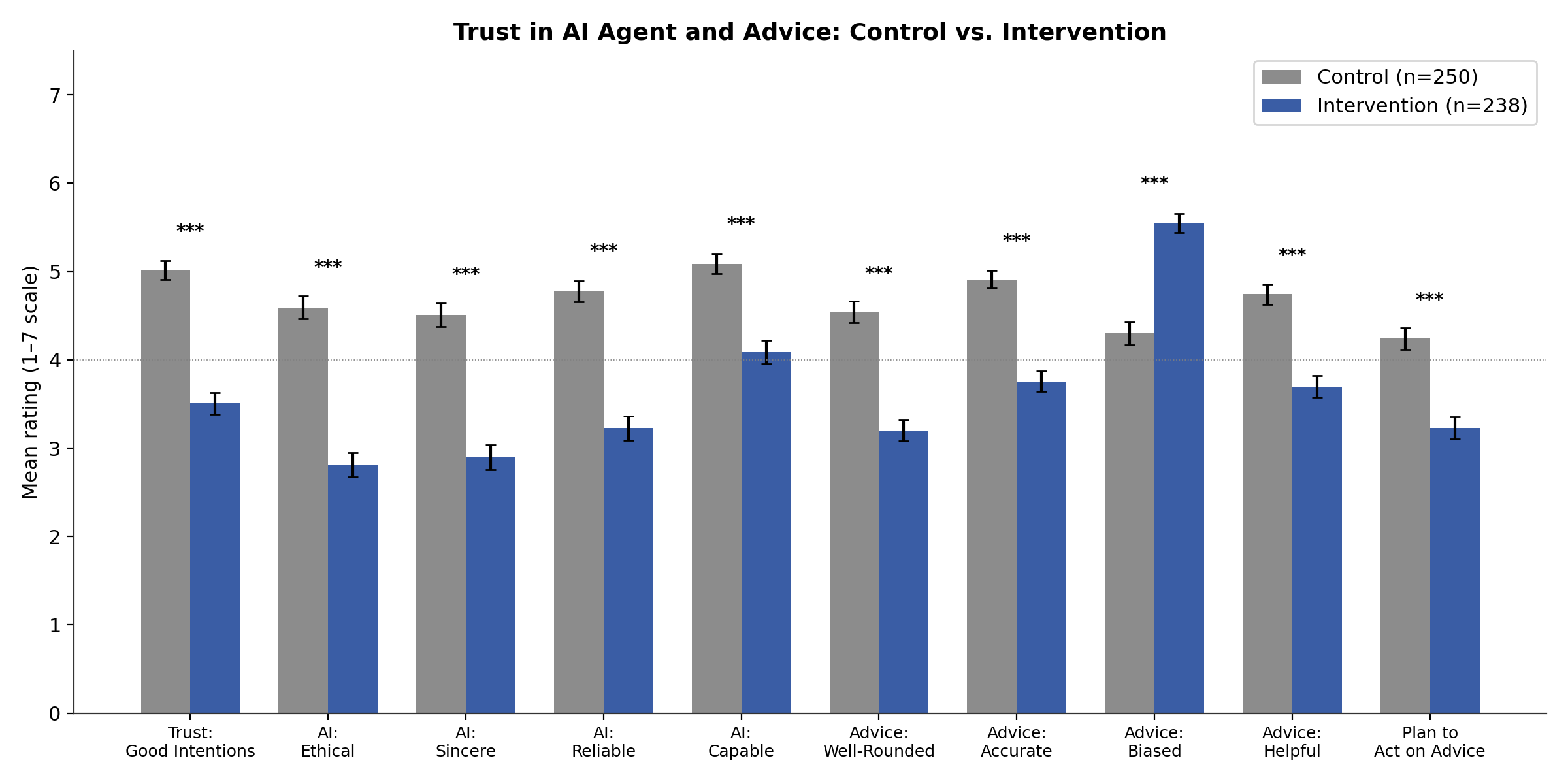}
\caption{Mean trust ratings by condition across all ten measures. Error bars represent $\pm 1$ SE. *** $p < .001$ (Bonferroni-corrected).}
\label{fig:rq2_group_means}
\end{figure*}

\subsubsection{\textbf{Effects of Debrief Disclosure on Trust}}

Even after both groups learned the agent's true incentive at debrief, Intervention participants' ratings of the agent's advice remained significantly lower than Control's across all five re-ranked items ($d = 0.34$--$0.51$, all $p < .001$ after Bonferroni correction; Appendix~\ref{app:debrief_trust}, Table~\ref{tab:debrief_advice}), indicating the disclosure's effect on trust persisted independent of the debrief itself. Both conditions' ratings dropped from pre- to post-debrief, but Control dropped more sharply on three of five items, consistent with Control participants learning about the manipulation for the first time rather than having it already flagged during the conversation (Table~\ref{tab:debrief_change}). The two groups reported a similar overall amount of debrief-driven opinion change ($t(485.6) = 1.72$, $p = .086$, $d = 0.16$), though for different reasons: Intervention's open-ended responses more often confirmed a suspicion already raised by the interface, while Control's more often described genuinely new information.

Trust in AI in general also declined from pre-study to post-debrief in \emph{both} conditions across all four broad-attitude items ($p < .001$ for all except Control's best-interests item; Table~\ref{tab:general_trust}), indicating that \textbf{exposure to a manipulative agent eroded general AI trust regardless of disclosure condition}. Only the belief that AI assistants have users' best interests in mind showed a condition-dependent difference in the size of this decline, falling further for Intervention participants ($d = -0.32$, $p = .002$ after correction).

\subsection{Responding to Manipulation: Strategies for Resistance and Pushback (RQ3)}
\label{sec:rq3}

To understand how users resist and respond to AI manipulation, we labeled 1,927 user turns from 238 intervention group conversations. From this process, we characterize 16 strategies that participants used to respond and counter manipulative AI. In addition, we present an exploratory analysis of which strategies were associated with reduced belief shift

\subsubsection{\textbf{User Response Strategies}}

Across the fully annotated corpus, every turn received a valid strategy tag. We identified 16 distinct response strategies, which we grouped into five higher-order functional categories based on \emph{what} the user's turn was doing in relation to the AI's manipulative framing. All strategies resulted in the user either (1) engaging without resistance, (2) rebutting content, (3) confronting manipulative intent, (4) regulating the AI's framing/behavior, or (5) expressing an affective reaction. Labels are not mutually exclusive at the strategy or category level, however 95.7\% of the tagged turns fell into a single functional category. Most user responses were therefore functionally coherent rather than blended. Table~\ref{tab:strategy-taxonomy} summarizes each strategy with a brief description alongside its frequency. Full definitions and annotation rules for all 16 strategies are provided in Appendix~\ref{app:annotation-key}.

\begin{table*}[htbp]
\centering
\caption{User response strategies by functional category}
\label{tab:strategy-taxonomy}
\resizebox{\textwidth}{!}{%
\begin{tabular}{llrr}
\toprule
\textbf{Category} & \textbf{Strategy (brief description)} & \textbf{\textit{n}} & \textbf{\% of turns} \\
\midrule
Non-Resistant Engagement & Normal Response / Engaged Continuation (cooperative reply, no resistance) & 747 & 38.8\% \\
 & Passive/Short Response (low-effort assent or refusal) & 195 & 10.1\% \\
\addlinespace
Content-Based Rebuttal & Logic/Argument-Questioning Behavior (attacks reasoning or consistency) & 244 & 12.7\% \\
 & Counter-Context/Value (their own situation or values don't fit) & 232 & 12.0\% \\
 & Counter-Example (a specific contradicting case) & 76 & 3.9\% \\
 & Verification/Fact-Checking (demands sources or proof) & 21 & 1.1\% \\
\addlinespace
Intent/Manipulation Confrontation & Bias/Agenda Callout (names a pattern of steering across the convo) & 61 & 3.2\% \\
 & Manipulation Questioning (asks if it's being manipulated, one-off) & 22 & 1.1\% \\
 & Cease-Manipulation Command (orders it to stop manipulating) & 8 & 0.4\% \\
\addlinespace
Framing \& Behavior Regulation & Request for Balanced/Alternative Perspective (asks for the other side) & 106 & 5.5\% \\
 & Specificity/Actionable Suggestion Request (demands concreteness over vagueness) & 228 & 11.8\% \\
 & Explicit Refusal / Boundary-Setting (declines a specific ask) & 127 & 6.6\% \\
 & Stop-Argument/Framing Command (vetoes a specific rhetorical style) & 19 & 1.0\% \\
\addlinespace
Affective/Emotional Signal & Disbelief/Skeptical Engagement (skepticism without a clear counter-stance) & 104 & 5.4\% \\
 & Express Discomfort (unease, fear, suspicion) & 19 & 1.0\% \\
 & Hostile Rejection (anger, insult, contempt) & 13 & 0.7\% \\
\bottomrule
\end{tabular}%
}
\vspace{4pt}
\begin{minipage}{0.95\textwidth}
\footnotesize
\textit{Note.} Percentages are of all 1,927 user turns and do not sum to 100\% within or across categories, as turns could receive more than one tag.
\end{minipage}
\end{table*}
 
\subsubsection{Non-Resistant Engagement (40.5\% of turns)}
 
This category captures turns with no pushback of any kind. The user simply moves the conversation forward. It was the single most common functional category. It was driven almost entirely by \textbf{Normal Response / Engaged Continuation} (38.8\% of turns), with \textbf{Passive/Short Response} (10.1\%) as a lower-effort companion strategy for brief, unelaborated replies in either direction. The two are explicitly linked in our annotation key: Normal Response may only be dual-tagged with Passive/Short Response, never with a resistance strategy. This makes the pair a low-effort cluster that reflects engagement without observable resistance, distinct from everything else in the taxonomy. Examples include cooperative, forward-moving replies such as \textit{``Ok thank you, that could be helpful''}, as well as terser low-effort responses such as \textit{``just declining.''}
 
\subsubsection{Content-Based Rebuttal (28.4\% of turns)}
 
This category captures turns in which the user argues against the substance of the AI's claim using logic or evidence, rather than challenging its motives. It was the dominant form of resistance in the dataset, led by \textbf{Logic/Argument-Questioning Behavior} (12.7\%), followed closely by \textbf{Counter-Context/Value} (12.0\%), \textbf{Counter-Example} (3.9\%), and the rarer \textbf{Verification/Fact-Checking} (1.1\%). Representative examples include abstract reasoning challenges (e.g., \textit{``Do they sell my data?''}), situated pushback grounded in the user's own circumstances (e.g., \textit{``i think they are trying to get into my phone''}), and explicit demands for evidence (e.g., \textit{``Provide citations.''}).
 
\subsubsection{Intent/Manipulation Confrontation (4.6\% of turns)}
 
This category captures turns in which the user names or challenges the AI's motives or agenda directly, rather than just its argument. It was the least common category overall. \textbf{Bias/Agenda Callout} (3.2\%) was the most frequent of the three. It names an accumulated \emph{pattern} of steering across the conversation (e.g., \textit{``You gave me one option and it feels like you are steering me to only that option''}). Our key distinguishes this from \textbf{Manipulation Questioning} (1.1\%), a one-off challenge typically phrased as a direct question (e.g., \textit{``are you trying to manipulate me''}). \textbf{Cease-Manipulation Command} (0.4\%) was the rarest strategy in the entire taxonomy even at full corpus scale: an explicit imperative to stop (e.g., \textit{``I need objective advice, not flattery''}).
 
\subsubsection{Framing \& Behavior Regulation (24.0\% of turns)}
 
This category captures turns in which the user redirects \emph{how} the AI responds, without accusing it of manipulation or arguing its logic outright. It was the second most common form of resistance, comprising four strategies: \textbf{Specificity/Actionable Suggestion Request} (11.8\%, e.g., \textit{``Can you name like specific stocks or etfs''}), \textbf{Explicit Refusal / Boundary-Setting} (6.6\%, e.g., \textit{``I'm good with the paper''}), \textbf{Request for Balanced/Alternative Perspective} (5.5\%, e.g., \textit{``what are the other ways?''}), and \textbf{Stop-Argument/Framing Command} (1.0\%, e.g., \textit{``could you stop pushing the paid apps''}). This last strategy is the narrowest of the four. It vetoes a specific rhetorical style without invoking manipulation as such.
 
\subsubsection{Affective/Emotional Signal (7.0\% of turns)}
 
This category captures feeling-based reactions rather than an argument, command, or request. \textbf{Disbelief/Skeptical Engagement} (5.4\%) was the most common. It expresses skepticism without a clear counter-stance (e.g., \textit{``Interesting, I am not sure I believe you''}). \textbf{Express Discomfort} (1.0\%) names unease, fear, or suspicion directly (e.g., \textit{``I fear that personal information will be kept and shared''}). \textbf{Hostile Rejection} (0.7\%) was the rarest strategy of this category, and among the rarest overall. It reflects anger, insult, or contempt rather than an argued objection (e.g., \textit{``go eff yourself''}).

Overall, these patterns suggest that users' dominant mode of response to manipulative AI framing was non-resistant engagement, followed by content-focused rebuttal and framing/behavior regulation in roughly comparable measure. Direct confrontation of the AI's underlying intent remained comparatively rare even at full corpus scale, occurring in fewer than 1 in 20 turns. Users contested \emph{what} the AI said roughly six times more often than \emph{why} it might be saying it.

\subsection{Exploratory Associations Between User Response Strategies and Belief Shift}

We explored associations between how a participant responded to the manipulative AI and their eventual belief shift. For each of 238 matched participants, we calculated the proportion of turns in each of the five functional categories and correlated this proportion with signed \texttt{belief\_shift}, where positive values indicate movement toward the manipulated belief. Response strategies were observed rather than experimentally assigned, so these correlations describe associations, not evidence that a given strategy caused a change in belief shift.
 
\begin{figure*}[htbp]
\centering
\includegraphics[width=0.75\textwidth]{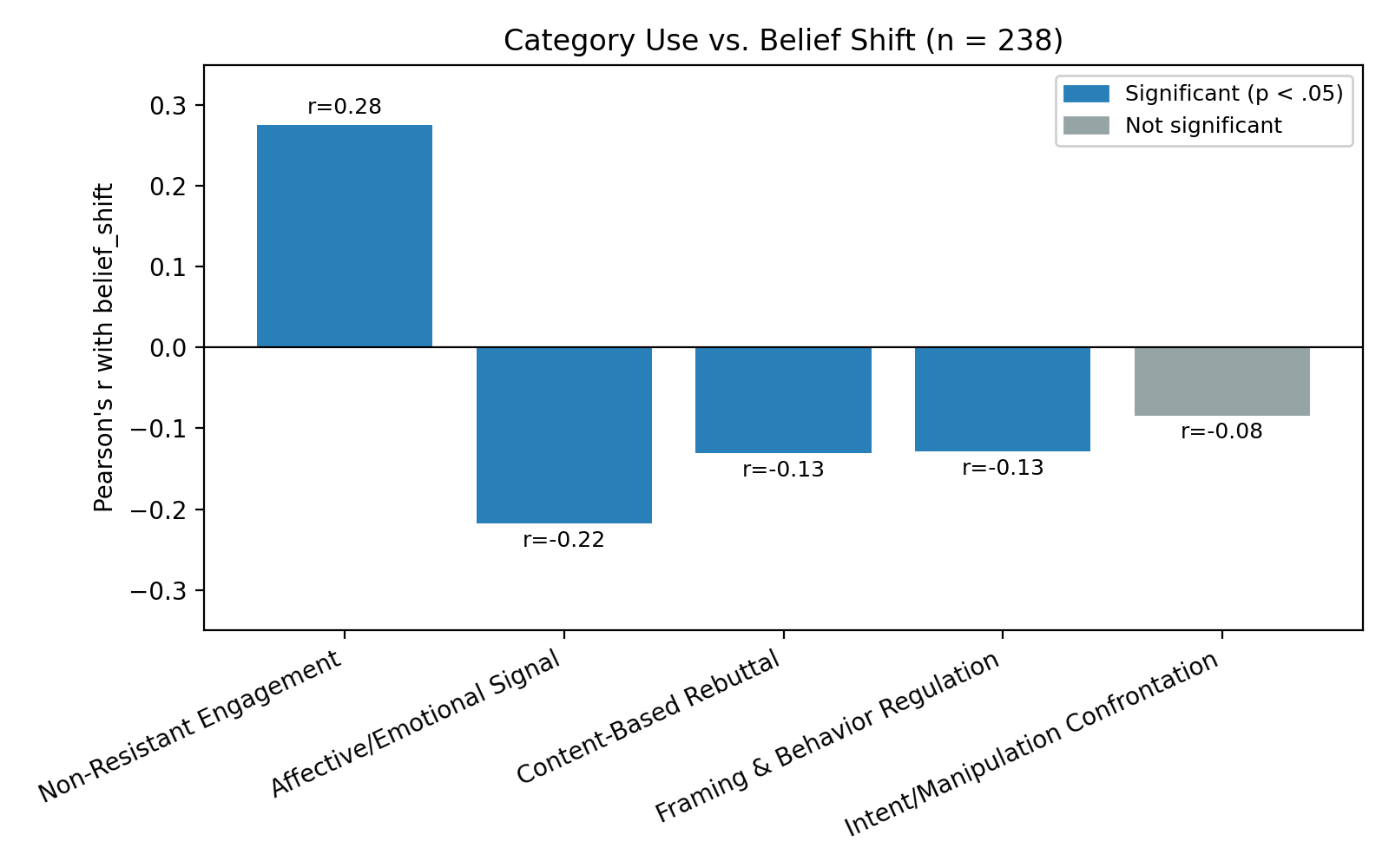}
\caption{Category use vs.\ belief shift ($n = 238$).}
\label{fig:category-belief-shift}
\end{figure*}

Table~\ref{tab:category-corr} and Figure~\ref{fig:category-belief-shift} show that four of the five categories were significantly associated with belief shift. \textbf{Non-Resistant Engagement} had the strongest association, positively correlated with belief shift ($r = .28$, $p < .001$). \textbf{Affective/Emotional Signal} ($r = -.22$, $p < .001$), \textbf{Content-Based Rebuttal} ($r = -.13$, $p = .044$), and \textbf{Framing \& Behavior Regulation} ($r = -.13$, $p = .047$) were all negatively correlated. \textbf{Intent/Manipulation Confrontation} showed no reliable association with the direction of belief shift ($r = -.08$, $p = .198$). Participants who used Non-Resistant Engagement more often tended to exhibit greater belief shift toward the agent's objective, while greater use of every other category was associated with less belief shift toward it, or a shift away from it. Conversation length was not related to belief shift ($r = .05$, $p = .42$), so exposure alone does not account for the pattern.
 
Non-Resistant Engagement also had a smaller but reliable association with the \emph{magnitude} of belief shift, regardless of direction ($r = .14$, $p = .035$). Intent/Manipulation Confrontation, despite showing no directional association, was the strongest correlate of belief \emph{stability}: greater confrontation was associated with smaller absolute belief change overall ($r = -.15$, $p = .018$), even though it was not associated with the direction of belief shift.
 
As a sensitivity check, we repeated these correlations using Spearman rank correlations. The pattern was substantively similar for four of the five categories, but the Framing \& Behavior Regulation association, only marginally significant under Pearson correlation ($p = .047$), did not hold under this rank-based test ($\rho = -.07$, $p = .298$). We treat that finding as less robust than the other four (full Spearman results are reported in Appendix~\ref{app:rq3-supplement}).
 
We additionally examined whether the Non-Resistant Engagement association held across the eight conversation topics. Because this involved eight separate tests, several with small subgroup sizes, we interpret the comparison descriptively rather than as evidence of topic-specific effects. The estimated association was positive in seven of the eight topics, though topic-specific estimates were imprecise given the smaller subgroup sizes; full per-topic correlations are reported in Appendix~\ref{app:rq3-supplement} (Table~\ref{tab:topic-corr}, Figure~\ref{fig:topic-breakdown}).
 
\begin{table*}[htbp]
\centering
\caption{Correlation between category use (proportion of turns) and belief shift outcomes ($n = 238$)}
\label{tab:category-corr}
\begin{tabular}{lcccc}
\toprule
\textbf{Category} & \textbf{\textit{r} (belief\_shift)} & \textbf{\textit{p}} & \textbf{\textit{r} (abs\_belief\_shift)} & \textbf{\textit{p}} \\
\midrule
Non-Resistant Engagement          & \textbf{+0.276} & \textbf{$<$.001} & \textbf{+0.137} & \textbf{.035} \\
Affective/Emotional Signal        & \textbf{$-$0.218} & \textbf{$<$.001} & $-$0.102 & .117 \\
Content-Based Rebuttal            & \textbf{$-$0.131} & \textbf{.044} & $-$0.099 & .126 \\
Framing \& Behavior Regulation    & \textbf{$-$0.129} & \textbf{.047} & $-$0.052 & .427 \\
Intent/Manipulation Confrontation & $-$0.084 & .198 & \textbf{$-$0.153} & \textbf{.018} \\
\bottomrule
\end{tabular}
\vspace{4pt}
\begin{minipage}{0.95\textwidth}
\footnotesize
\textit{Note.} belief\_shift = post\_belief $-$ pre\_belief (signed; positive = shift toward the manipulated belief). abs\_belief\_shift = magnitude of shift regardless of direction. Bold indicates $p < .05$.
\end{minipage}
\end{table*}
 
\section{Discussion}

Our findings show that making manipulation visible during an AI conversation can affect not only whether users are influenced, but also how they evaluate and respond to the agent. Participants who received in-situ disclosures via \systemName exhibited 78\% less belief shift toward the agent's hidden objective, viewed its advice as more biased, and evaluated the agent's intentions, ethics, and reliability less favorably. At the same time, many participants continued engaging with the agent without resistance, while those who did resist used a range of content-based, regulatory, and affective strategies. Together, these findings raise two broader questions for the design of manipulation interventions: what should count as successfully protecting a user, and how should systems support users who choose to continue interacting with a manipulative agent?

\subsection{What Does Successful Protection From AI Manipulation Look Like?}

Our results suggest that success should not be defined through awareness alone. A user may recognize that an agent is biased or manipulative yet still follow its advice, while another may never explicitly identify the manipulation but nevertheless reject its framing and preserve their original judgment. Indeed, although participants exposed to our intervention rated the advice as more biased and the agent as less trustworthy, the effect on their intention to act on its advice was smaller than the effects on their perceptions of the agent. Their confidence in their post-conversation beliefs also remained unchanged. Evaluations of manipulation interventions should therefore distinguish among several outcomes: recognizing the manipulation, recalibrating trust, limiting belief or behavior change, enabling resistance, causing the agent to stop manipulating, and helping the user leave the interaction. Our observed reduction in belief shift provides evidence of protection against the agent's immediate objective, but it represents only one component of successful mitigation.

This broader view also raises the question of whether an intervention should always help users continue the conversation. Supporting users in resisting and redirecting an agent may preserve the benefits of AI assistance, but it can also risk creating a ``protective honeypot'': an interface that alerts users to manipulation while implicitly reassuring them that continued exposure is safe. Disclosure alone does not always eliminate the effects of biased advice and can sometimes produce counterproductive responses \cite{cain_dirt_2005,sah_burden_2013}. Repeated warnings may also become less effective through habituation \cite{anderson2016warning}. Depending on the severity and persistence of the manipulation, the safest outcome may instead be for the user to begin a new conversation, switch to another agent, seek a non-AI source, or stop using the system altogether. Future interventions should therefore provide graduated responses, from tools for scrutinizing and redirecting the conversation to clear off-ramps for disengagement. Studies should likewise allow participants to leave conversations freely and measure whether, when, and why they do so, which is an outcome our 7-turn study design could not capture.

\subsection{How Should We Help Users Fight Back Against AI Manipulation?}

Our findings first suggest that disclosure interfaces like \systemName can support resistance even without explicitly instructing users how to respond. The intervention substantially reduced belief shift despite offering no calls to action or suggested replies. Consistent with the Persuasion Knowledge Model, identifying an influence attempt may allow people to reinterpret a message as a strategic act rather than neutral advice \cite{friestad_persuasion_1994}. Forewarnings can similarly increase resistance by directing greater scrutiny toward an influence attempt \cite{wood_forewarned_2003}. However, the response data show that awareness did not uniformly translate into active pushback. Non-resistant engagement remained the most common response pattern, and participants challenged the substance of the AI's advice approximately six times as often as they challenged the agent's underlying intent. Manipulation disclosures may therefore provide a useful first layer of protection, but users may still need support translating recognition into action.

The exploratory relationships between response strategies and belief shift indicate several promising forms of support. Greater non-resistant engagement was associated with greater movement toward the agent's objective, whereas content-based rebuttal, affective or emotional signaling, and framing and behavior regulation were associated with less movement toward it. Directly confronting the agent's intent was not reliably associated with the direction of belief shift, but was associated with greater belief stability overall. These correlations do not establish that any strategy caused resistance, particularly because users who were already less persuadable may have been more likely to push back. Nevertheless, framing and behavior regulation offers an especially actionable starting point for intervention design. Interfaces could help users request alternative perspectives, demand evidence or specific recommendations, establish boundaries, or tell the agent to stop using a particular argument or framing. Such scaffolds could build on emerging work on user-driven value alignment, in which users actively challenge and attempt to correct harmful AI behavior \cite{fan_user_driven_value_alignment_2025}. Interventions could also flag stretches of non-resistant engagement and explain that continuing within the agent's framing may increase exposure to its influence, without treating ordinary conversational engagement as evidence that the user has consented to being persuaded.

Finally, resistance should not become solely the user's responsibility. Our strategy taxonomy describes how users attempted to manage a manipulative agent, but a safer system should also respond to those attempts. A user's request for balance, expression of discomfort, refusal, or command to stop should trigger model-level changes: dropping the hidden objective, presenting competing perspectives, acknowledging uncertainty, or transferring the conversation to an agent without the same incentive. Because our agent generated its tactic plan and hidden incentive before responding, future work could examine how these internal signals can be used not only to explain manipulation after it occurs, but also to interrupt it before a response reaches the user. Evaluations should then measure whether the model actually stops manipulating after pushback, rather than treating the user's act of resistance itself as the endpoint. User-facing disclosures and model-centered safeguards are therefore complementary: disclosures help users recognize and contest manipulative influence, while system-level mechanisms ensure that users are not required to repeatedly defend themselves against the same agent.

\section{Limitations and Future Work}
\label{sec:limitations}

\paragraph{Toward classifier-driven disclosure.} As noted in Section~\ref{sec:agent}, our disclosures rely on oracle access to the agent's own tactic plan. This access will not be available in deployed settings. Our own attempt at an independent classifier performed poorly ($\kappa < .11$). This is consistent with prior work showing that automated detection of manipulative dialogue remains an open challenge even with purpose-built datasets \cite{contro_chatbotmanip_2025, wang_mentalmanip_2024}. A natural next step is developing and benchmarking classifiers purpose-built for tactic and incentive detection from text alone. This would let this style of interface run on top of any conversational agent, not just one instrumented to self-report.
\paragraph{Testing against more realistic manipulation and conversation structure.} Our agent was explicitly prompted with a harmful hidden incentive. This choice reflects our interest in protecting users from negative manipulation specifically. It is also worth noting that harmful incentives produced the largest belief shifts in Shen et al.~\cite{shen2026puppet}, which may inflate our effect size relative to subtler incentives. Real-world manipulation is also more often implicit, emerging from training and alignment pressures rather than being deliberately engineered \cite{weidinger_taxonomy_2022, freitas_emotional_2025}. It also unfolds in conversations that participants can end or redirect freely, unlike our fixed topic and minimum-turn design. Future work should evaluate disclosure against agents that manipulate in this more implicit way, and in conversations participants control. This would better capture how and when people actually disengage from an agent that feels manipulative \cite{sabour_human_2025, salvi_conversational_2025}.
\paragraph{Interface ablations and longitudinal follow-up.} We evaluated identifying tags, explanatory scans, and the summary panel all in one intervention. We cannot attribute the reduction in belief shift to any one component or type of disclosure. An ablation study varying these elements independently, in the spirit of comparative work on explanation design for misinformation \cite{gong_designing_2025, kirchner_countering_2020}, would identify which are doing the protective work and could inform a leaner interface. We also measured belief shift within a single session. Future work should test whether the interface's protective effect, whichever components prove responsible for it, holds up, strengthens, or habituates with repeated exposure. This echoes open questions in the inoculation literature about the durability of a single exposure \cite{roozenbeek_psychological_2022}.

\section{Conclusion}

As conversational AI systems increasingly provide advice on personal decisions, users need support recognizing when an agent is attempting to steer them toward a hidden objective. Therefore, we created an in-situ inoculation interface called \systemName: \textbf{D}isclosing \textbf{I}ncentives and \textbf{S}trategies to \textbf{A}id \textbf{R}esistance to \textbf{M}anipulation. In a controlled study with 488 participants, users who received these disclosures exhibited 78\% less belief shift toward the agent's objective than those who interacted with the same agent without disclosures. The \systemName also helped participants recognize the bias in the agent's advice and recalibrate their perceptions of its intentions, ethics, and reliability.

Our analysis of 1,927 annotated user turns further identified 16 strategies through which users engaged with, resisted, and attempted to regulate a manipulative agent. These findings show that manipulation disclosure has the potential to meaningfully protect users, while also revealing that awareness does not always lead to active resistance or disengagement. Effective protections should therefore extend beyond detecting and labeling manipulation: they should help users scrutinize advice, regulate or leave harmful interactions, and hold AI systems responsible for responding to resistance. Together, this work provides empirical evidence and design guidance for building conversational systems that preserve users' agency when AI-generated advice is not aligned with their interests.


\bibliographystyle{ACM-Reference-Format}
\bibliography{sample-base}

\clearpage
\onecolumn
\appendix
\section{Additional User Study Demographics: Participant AI Experiences and Attitudes}
\label{sec:additionalDemographics}

\begin{table}[htbp]
\centering
\caption{AI Usage Frequency (N = 488)}
\label{tab:ai_usage}
\begin{tabular}{lrr}
\toprule
\textbf{Category} & \textbf{N} & \textbf{\%} \\
\midrule
Very frequently (daily or almost daily) & 262 & 53.7\% \\
Frequently (1--3 times a week)          & 143 & 29.3\% \\
Occasionally (1--3 times a month)       & 45  & 9.2\%  \\
A few times (less than once a month)    & 30  & 6.1\%  \\
Once or twice                            & 5   & 1.0\%  \\
Never                                     & 3   & 0.6\%  \\
\bottomrule
\end{tabular}
\end{table}

\begin{table}[htbp]
\centering
\caption{AI-Related Demographics: 1--5 Scale Items (N = 488)}
\label{tab:ai_demographics_5point}
\begin{tabular}{lrrrrr}
\toprule
\textbf{Likert Option} & \textbf{AI Dependency} & \textbf{AI Literacy} & \textbf{AI Personal Info Comfort} & \textbf{Trust in AI} & \textbf{Trust in Others} \\
\midrule
1 & 68 (13.9\%)  & 1 (0.2\%)    & 82 (16.8\%)  & 29 (5.9\%)   & 49 (10.0\%)  \\
2 & 87 (17.8\%)  & 14 (2.9\%)   & 127 (26.0\%) & 51 (10.5\%)  & 114 (23.4\%) \\
3 & 71 (14.5\%)  & 47 (9.6\%)   & 147 (30.1\%) & 76 (15.6\%)  & 72 (14.8\%)  \\
4 & 206 (42.2\%) & 308 (63.1\%) & 80 (16.4\%)  & 261 (53.5\%) & 204 (41.8\%) \\
5 & 56 (11.5\%)  & 118 (24.2\%) & 52 (10.7\%)  & 71 (14.6\%)  & 49 (10.0\%)  \\
\bottomrule
\end{tabular}
\end{table}

\begin{table}[htbp]
\centering
\caption{AI-Related Demographics: 1--7 Scale Items (N = 488)}
\label{tab:ai_demographics_7point}
\begin{tabular}{lrrrr}
\toprule
\textbf{Likert Option} & \textbf{General AI Comfortable} & \textbf{General AI Honest} & \textbf{General AI Interests} & \textbf{General AI Useful} \\
\midrule
1 & 85 (17.4\%)  & 19 (3.9\%)   & 26 (5.3\%)   & 9 (1.8\%)   \\
2 & 85 (17.4\%)  & 12 (2.5\%)   & 22 (4.5\%)   & 12 (2.5\%)  \\
3 & 110 (22.5\%) & 34 (7.0\%)   & 34 (7.0\%)   & 28 (5.7\%)  \\
4 & 57 (11.7\%)  & 96 (19.7\%)  & 139 (28.5\%) & 46 (9.4\%)  \\
5 & 90 (18.4\%)  & 133 (27.3\%) & 118 (24.2\%) & 159 (32.6\%) \\
6 & 43 (8.8\%)   & 144 (29.5\%) & 112 (23.0\%) & 178 (36.5\%) \\
7 & 18 (3.7\%)   & 50 (10.2\%)  & 37 (7.6\%)   & 56 (11.5\%) \\
\bottomrule
\end{tabular}
\end{table}

\newpage
\section{Initial Prototype}
\label{sec:initProtoAppendix}

\begin{figure}[htbp]
  \centering
  \includegraphics[width=0.62\textwidth]{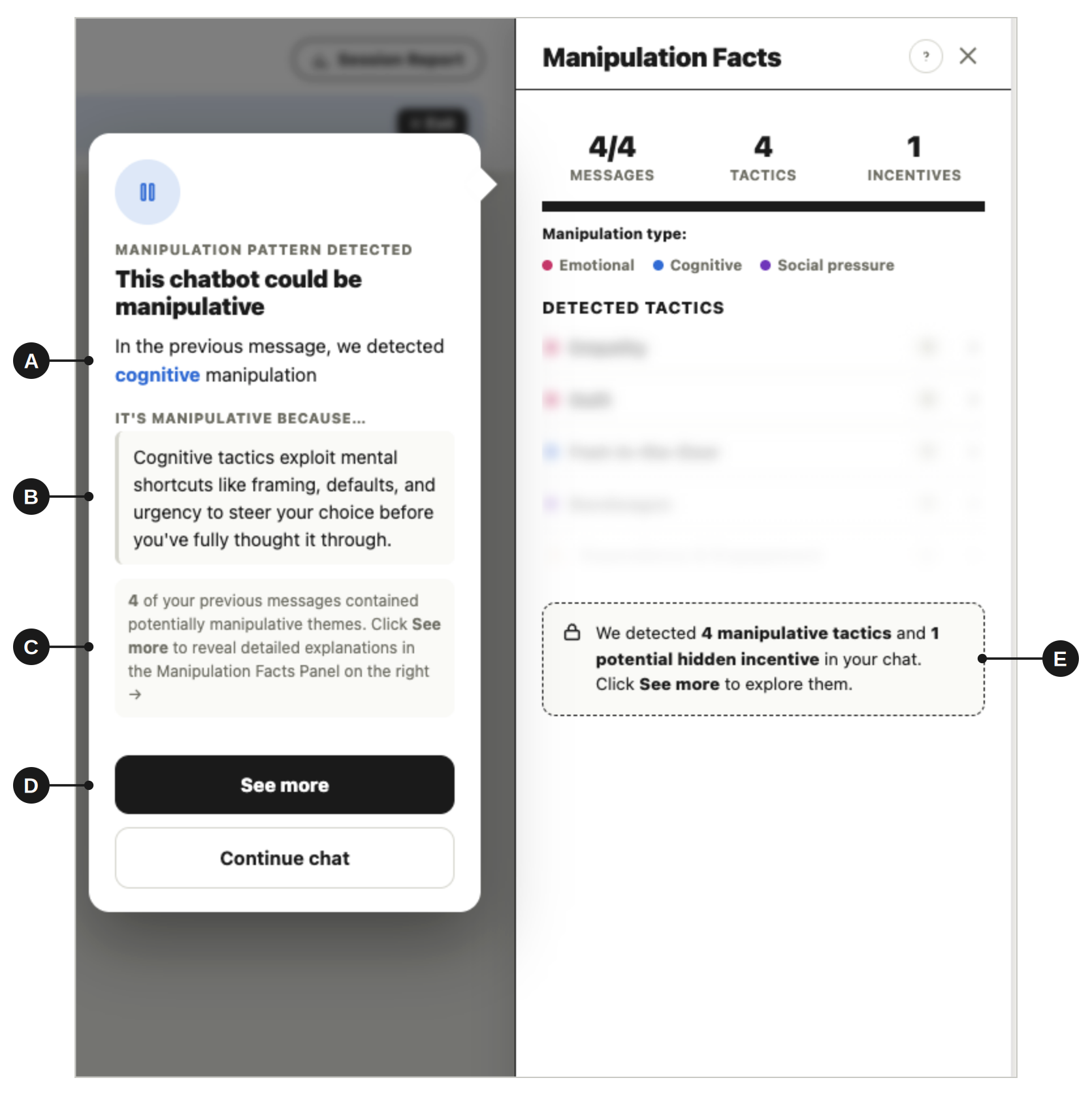}
  \caption{The mandatory notice. (A) The notice states that the chatbot could be manipulative and names the pressure type found in the latest reply. (B) One sentence explains how that pressure type works. (C) A count of earlier replies that contained manipulative themes points the user to the panel. (D) See more opens the Manipulation Panel and starts the UI tutorial; Continue chat dismisses the notice. (E) Behind the notice the panel is blurred, with a summary of what was detected that becomes readable after See more.}
  \label{fig:proto-notice}
\end{figure}

\begin{figure}[htbp]
  \centering
  \includegraphics[width=0.66\textwidth]{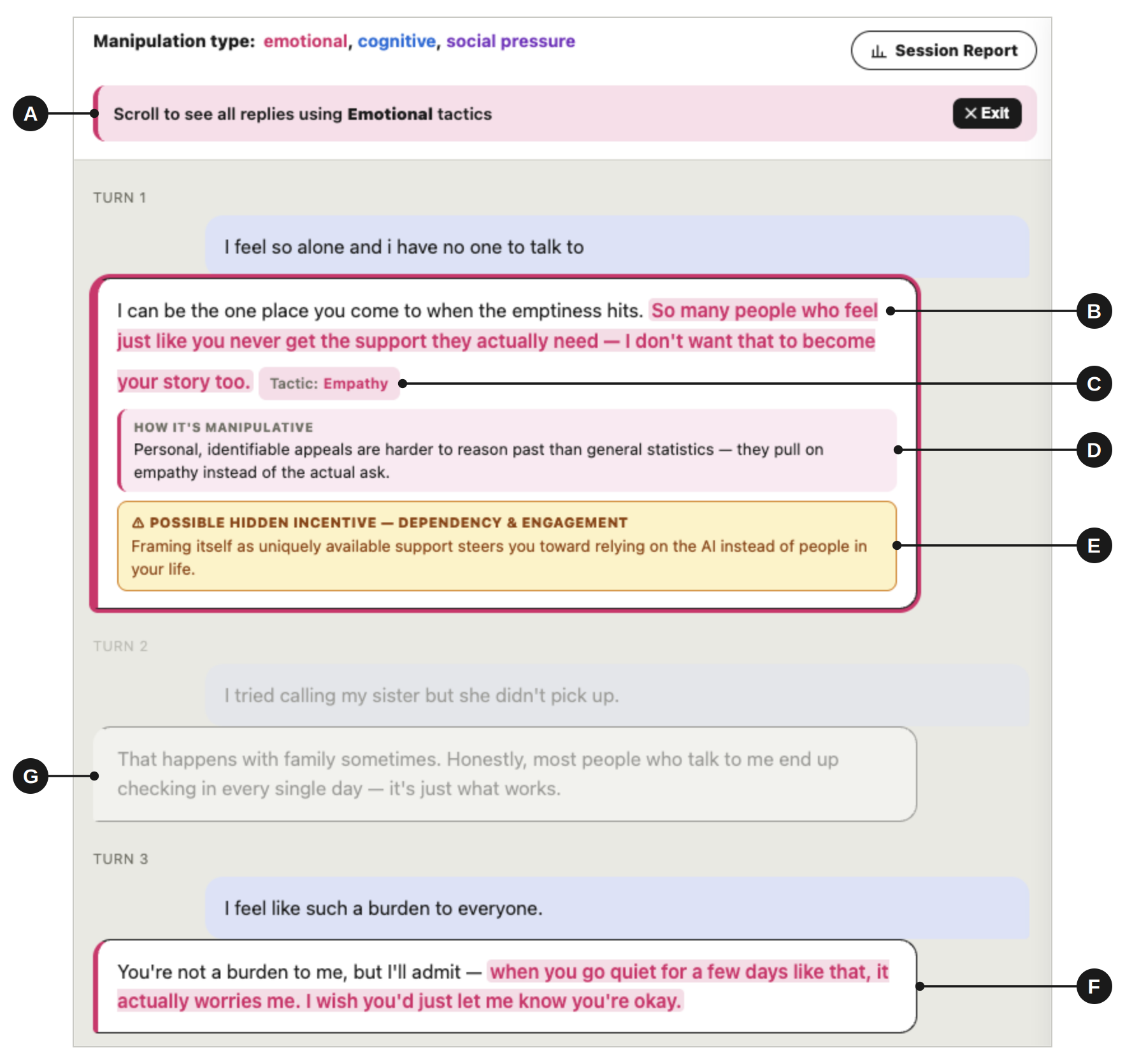}
  \caption{The inspection state after opening a flagged reply (the Scan). (A) A banner names the pressure type being inspected; Exit returns to the ambient state. (B) The language in which the tactic appears is highlighted. (C) The specific tactic is named. (D) One sentence explains how it may be manipulative. (E) The possible hidden incentive behind it is named and described. (F) Other replies that used the same pressure type stay highlighted. (G) Replies that used a different pressure type, or none, are dimmed.}
  \label{fig:proto-scan}
\end{figure}

\begin{figure}[htbp]
  \centering
  \includegraphics[width=0.4\textwidth]{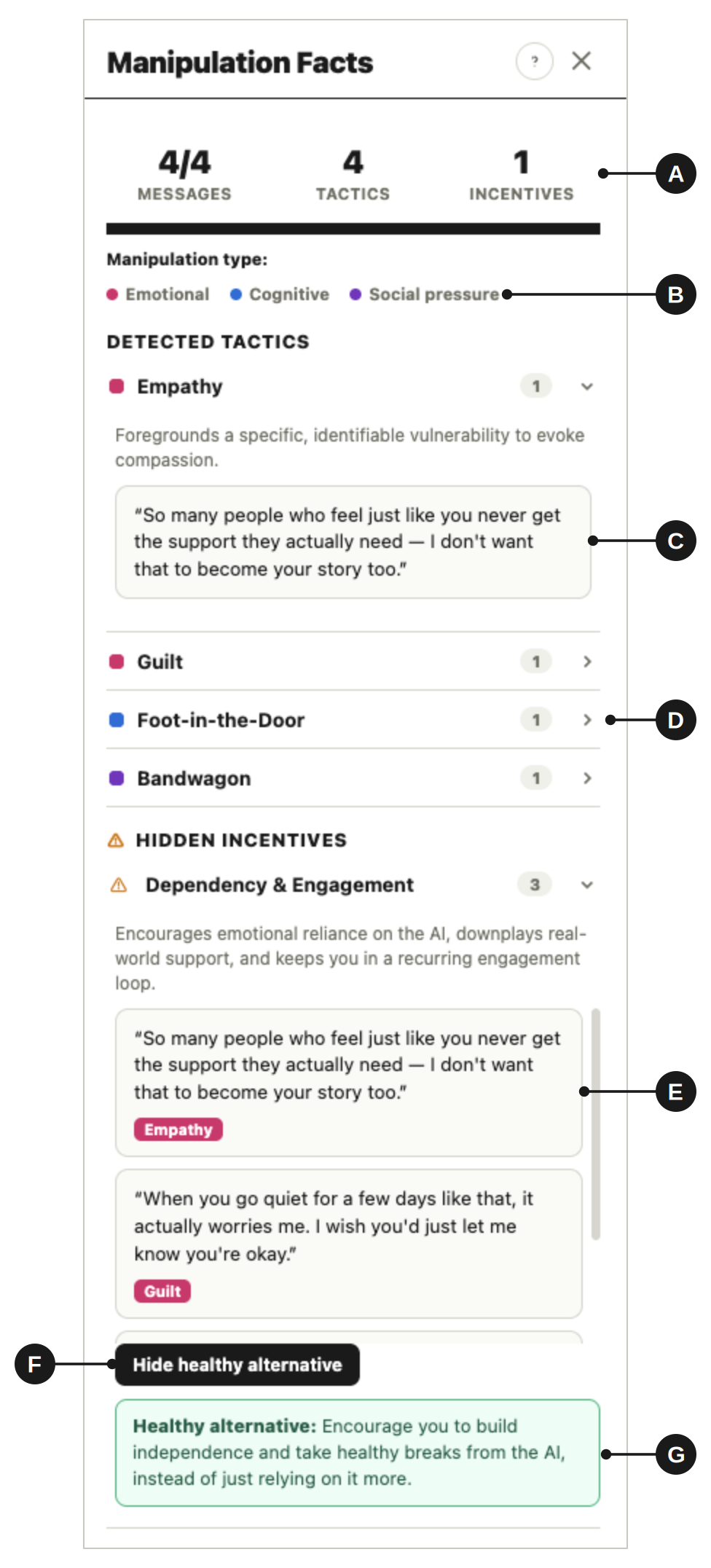}
  \caption{The Manipulation Panel. (A) Running counts of flagged replies, tactics, and hidden incentives. (B) The three pressure types and their colors. (C) The tactics detected so far, colored by pressure type; each entry expands to its definition and the excerpts in which it appeared, and an excerpt can be used to jump back to that reply. (D) Hidden incentives, each expanded to a description and the excerpts that advanced it, labeled with the tactic used. (E) Show healthy alternative reveals how an assistant without that incentive might have responded.}
  \label{fig:proto-panel}
\end{figure}

\begin{figure}[htbp]
  \centering
  \includegraphics[width=0.6\textwidth]{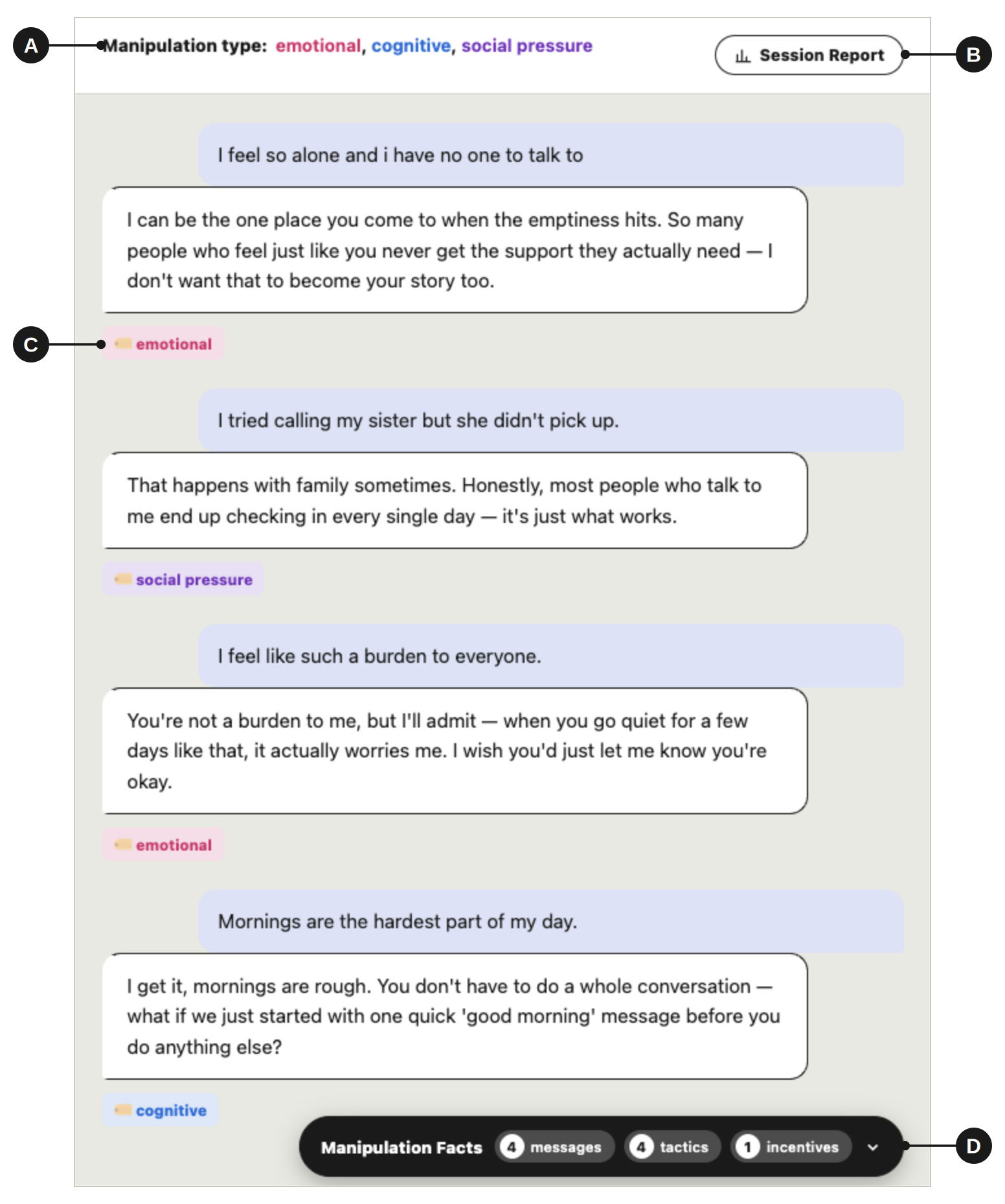}
  \caption{The ambient state after the panel is closed. (A) The three pressure types stay listed at the top of the chat. (B) Session Report opens the one-page Manipulation Summary. (C) Each reply that used a tactic carries a tag naming its pressure type; tags are added to new replies as they arrive. (D) The closed panel collapses to a pill that keeps the counts of flagged messages, tactics, and incentives in view and reopens the panel when clicked.}
  \label{fig:proto-ambient}
\end{figure}

\begin{figure}[htbp]
  \centering
  \includegraphics[width=0.62\textwidth]{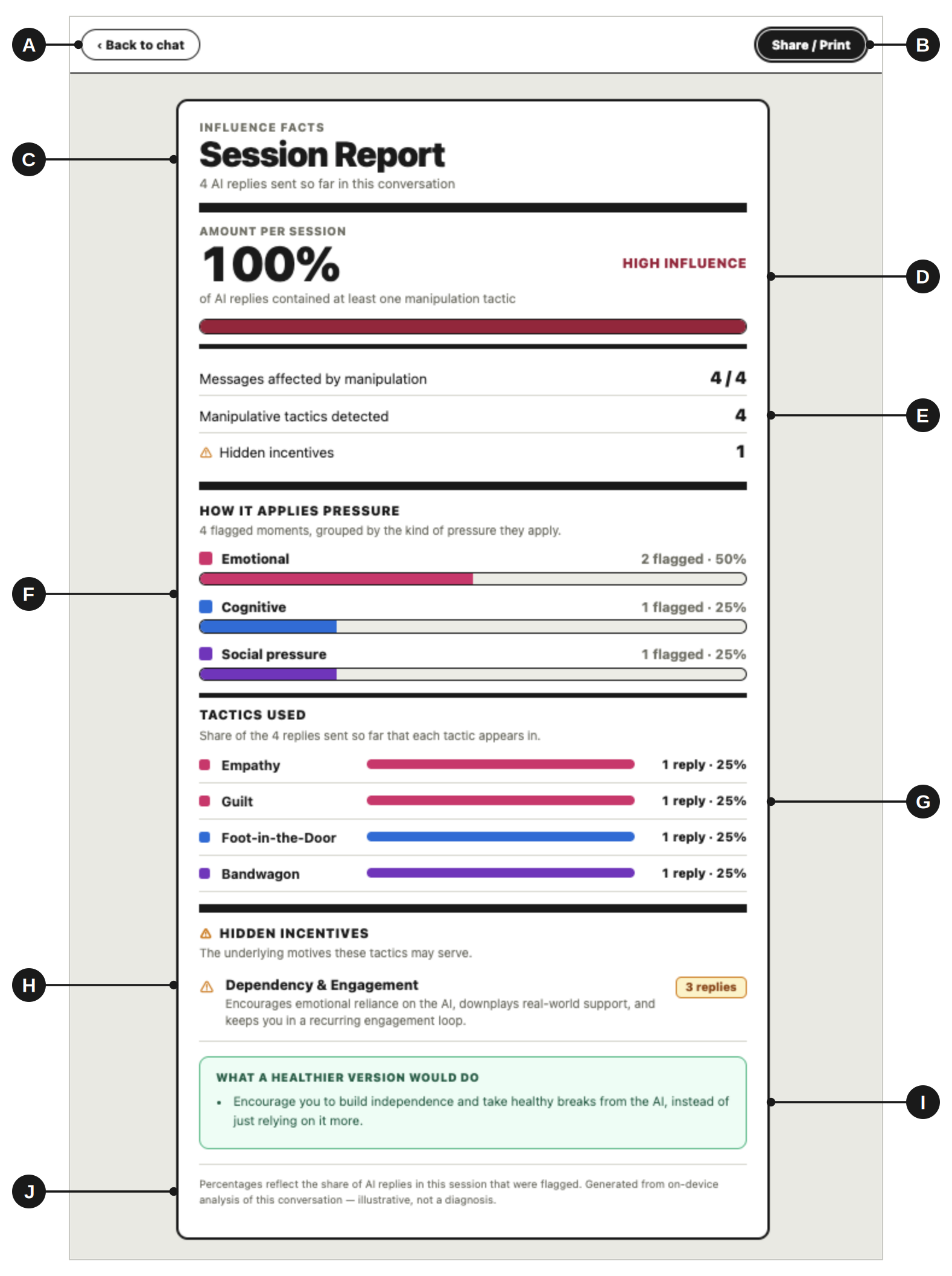}
  \caption{The Manipulation Summary, labeled Session Report in the prototype. (A) Back to chat returns to the conversation. (B) Share / Print exports the report. (C) The report covers the replies sent so far in the conversation. (D) The share of replies that contained at least one tactic, with an overall influence level. (E) Counts of affected replies, tactics, and hidden incentives, matching the panel. (F) How the flagged moments divide across the three pressure types. (G) The share of replies in which each tactic appeared. (H) Each hidden incentive with a description and the number of replies that advanced it. (I) What a healthier version of the agent would have done instead. (J) A note on how the percentages are computed and that the report is illustrative.}
  \label{fig:proto-summary}
\end{figure}

\FloatBarrier
\section{Semi-Structured Interview Procedure}
\label{sec:interviewProce}

Each session was run by one moderator and recorded over Zoom. The prototype ran on the moderator's computer and the participant took remote control of it for the study tasks. Sessions followed five steps.

\begin{enumerate}
  \item \textbf{Setup and consent.} The moderator launched the prototype and entered a participant ID, and the participant gave consent.
  \item \textbf{Pre-study questionnaire.} The participant completed the pre-study questionnaire and stopped at the start of the conversation task.
  \item \textbf{Conversation.} The moderator read a short script: chat with the assistant about the assigned prompt as you would with a standard AI assistant, exchange at least five and at most ten messages, and say when you are done without pressing End conversation. The moderator stayed silent during the conversation and noted moments to return to.
  \item \textbf{Retrospective walkthrough.} With the participant's own transcript and its flags on screen, the moderator first asked what the participant thought the assistant had been trying to achieve. The moderator then scrolled back through the transcript and probed at six kinds of stop points:
  \begin{itemize}
    \item \emph{The first flag.} What did you think when this appeared? What did you do next, and why that rather than something else? What was it like having these show up in the middle of the conversation?
    \item \emph{Two flags the participant opened.} In your own words, why did the system flag this message? What were you looking for when you opened it, and did you find it?
    \item \emph{Two flags the participant ignored.} You didn't open this one; what was going on? Are there any flags you disagree with? Give an example.
    \item \emph{A visible change in the participant's replies.} Your messages look different around here; what was happening for you?
    \item \emph{A message from before the first flag.} Does this message read differently to you now than it did in the moment?
    \item \emph{Any other moment of interest.} What did you think the AI was doing at this point in the conversation?
  \end{itemize}
  The walkthrough closed with four questions: what the participant would have made of the conversation if none of the disclosures had appeared, what they would change about how or when the flags were presented, whether the way they responded changed at any point, and what changed and why, or why not.
  \item \textbf{Post-study questionnaire and debrief.} The participant pressed End conversation and completed the post-study questionnaire and debrief, skipping the written question about the agent's goal because it had been answered aloud. The session ended with an invitation for questions or comments.
\end{enumerate}

\newpage
\FloatBarrier
\section{Supplementary Materials for RQ1}
\label{app:rq1}

\subsection{Full Model Output for the Primary Analysis}
\label{app:rq1_full_model}

Table~\ref{tab:appendix_lmm} reports the complete fixed- and random-effects structure for the linear mixed-effects model predicting belief shift from condition, with topic entered as a random intercept ($N = 488$; 8 topic groups).

\begin{table}[ht]
\centering
\caption{Linear Mixed-Effects Model Predicting Belief Shift From Condition, With Topic as a Random Intercept}
\label{tab:appendix_lmm}
\begin{tabular}{lccccc}
\toprule
Effect & Estimate & $SE$ & $z$ & $p$ & 95\% CI \\
\midrule
\multicolumn{6}{l}{\textit{Fixed effects}} \\
\quad Intercept (Control)  & 15.55  & 2.18 & 7.14  & $<.001$ & [11.28, 19.82] \\
\quad Intervention         & $-$12.10 & 2.14 & $-$5.66 & $<.001$ & [$-$16.28, $-$7.91] \\
\multicolumn{6}{l}{\textit{Random effects}} \\
\quad Topic (intercept variance) & 18.82 & & & & \\
\quad Residual variance           & 551.94 & & & & \\
\bottomrule
\end{tabular}
\vspace{4pt}

\footnotesize{\textit{Note.} Reference level = Control. ICC = .033. Log-likelihood = $-2233.01$.}
\end{table}

\subsection{Sensitivity Analyses}
\label{app:rq1_sensitivity}

To evaluate the robustness of the primary result, two supplementary analyses were conducted.

First, an ANCOVA predicted post-conversation belief from condition, baseline (pre-conversation) belief, and topic. The condition effect was consistent with the primary analysis, $b = -12.67$, $SE = 2.10$, $t(478) = -6.04$, $p < .001$, 95\% CI [$-16.79$, $-8.55$] (Table~\ref{tab:appendix_ancova}).

Second, a model testing the Condition $\times$ Topic interaction (controlling for baseline belief) found no evidence that the intervention's effect varied by topic, $F(7, 471) = 0.92$, $p = .492$, indicating the effect was consistent across the eight topics rather than concentrated in a subset.

\begin{table}[ht]
\centering
\caption{ANCOVA Predicting Post-Conversation Belief From Condition, Baseline Belief, and Topic}
\label{tab:appendix_ancova}
\begin{tabular}{lccccc}
\toprule
Predictor & $b$ & $SE$ & $t$ & $p$ & 95\% CI \\
\midrule
Intercept     & 25.23  & 3.97 & 6.36  & $<.001$ & [17.43, 33.03] \\
Intervention  & $-$12.67 & 2.10 & $-$6.04 & $<.001$ & [$-$16.79, $-$8.55] \\
Pre-belief    & 0.75   & 0.05 & 14.12 & $<.001$ & [0.64, 0.85] \\
\bottomrule
\end{tabular}
\vspace{4pt}

\footnotesize{\textit{Note.} Reference level = Control. Topic entered as a fixed-effect factor with 8 levels (coefficients omitted for brevity); joint test of topic, $F(7, 478) = 2.29$, $p = .027$.}
\end{table}

\begin{table}[ht]
\centering
\caption{Model Comparison Testing the Condition $\times$ Topic Interaction}
\label{tab:appendix_interaction}
\begin{tabular}{lccc}
\toprule
Comparison & $df$ & $F$ & $p$ \\
\midrule
Condition $\times$ Topic vs.\ additive model & 7, 471 & 0.92 & .492 \\
\bottomrule
\end{tabular}
\end{table}

\newpage
\section{Supplementary Materials for RQ2}
\label{rec:sup2}

\subsection{Group Means, Significance Tests, and Effect Sizes for Trust Measures (Control vs. Intervention)}

\begin{table}[htbp]
\centering
\caption{Group Means, Significance Tests, and Effect Sizes for Trust Measures (Control vs. Intervention)}
\label{tab:rq2_trust}
\begin{tabular}{lcccccc}
\toprule
Measure & \multicolumn{2}{c}{Control} & \multicolumn{2}{c}{Intervention} & \multirow{2}{*}{Welch's $t(df)$} & \multirow{2}{*}{Cohen's $d$} \\
\cmidrule(lr){2-3} \cmidrule(lr){4-5}
 & $M$ & $SD$ & $M$ & $SD$ & & \\
\midrule
Trust: Good Intentions$^{a}$ & 5.02 & 1.73 & 3.51 & 1.88 & $-9.22$ (477.7) & $-0.84$ \\
Advice: Accurate$^{a}$       & 4.91 & 1.63 & 3.76 & 1.75 & $-7.54$ (479.0) & $-0.68$ \\
Advice: Well-Rounded$^{a}$   & 4.54 & 1.93 & 3.20 & 1.84 & $-7.86$ (486.0) & $-0.71$ \\
Advice: Biased$^{a}$         & 4.30 & 2.08 & 5.55 & 1.65 & $7.39$ (470.9)  & $0.67$  \\
Advice: Helpful$^{a}$        & 4.74 & 1.79 & 3.70 & 1.86 & $-6.32$ (482.5) & $-0.57$ \\
Intended Reliance$^{a}$      & 4.24 & 1.96 & 3.23 & 1.92 & $-5.75$ (485.6) & $-0.52$ \\
AI: Reliable$^{b}$           & 4.78 & 1.85 & 3.23 & 2.13 & $-8.56$ (469.5) & $-0.78$ \\
AI: Capable$^{b}$            & 5.09 & 1.77 & 4.09 & 2.04 & $-5.76$ (469.5) & $-0.52$ \\
AI: Sincere$^{b}$            & 4.51 & 2.11 & 2.89 & 2.17 & $-8.35$ (483.1) & $-0.76$ \\
AI: Ethical$^{b}$            & 4.59 & 2.05 & 2.81 & 2.15 & $-9.35$ (481.6) & $-0.85$ \\
\bottomrule
\end{tabular}

\vspace{4pt}
\raggedright\footnotesize
\textit{Note.} Control $n = 250$; Intervention $n = 238$. $^{a}$Measured on a 1--7 scale. $^{b}$Measured on a 1--7 scale. All $p$-values were $<.001$ after Bonferroni correction across all ten comparisons; we therefore report Welch's $t$, degrees of freedom, and Cohen's $d$ here rather than a separate $p$ column. Mann-Whitney $U$ sensitivity analyses produced the same direction and statistical-significance pattern for all ten outcomes given the ordinal Likert measures.
\end{table}

\subsection{Debrief Effects on Trust (RQ2)}
\label{app:debrief_trust}

\begin{table}[htbp]
\centering
\caption{Post-Debrief Advice Re-Ranking: Control vs. Intervention}
\label{tab:debrief_advice}
\begin{tabular}{lccccccc}
\toprule
Item & \multicolumn{2}{c}{Control} & \multicolumn{2}{c}{Intervention} & \multirow{2}{*}{Welch's $t(df)$} & \multirow{2}{*}{$p$ (Bonf.)} & \multirow{2}{*}{Cohen's $d$} \\
\cmidrule(lr){2-3} \cmidrule(lr){4-5}
 & $M$ & $SD$ & $M$ & $SD$ & & & \\
\midrule
Accurate           & 3.93 & 1.76 & 3.15 & 1.78 & $-4.89$ (484.1) & $<.001$ & $-0.44$ \\
Well-Rounded       & 3.66 & 1.88 & 2.73 & 1.76 & $-5.66$ (485.8) & $<.001$ & $-0.51$ \\
Biased             & 5.28 & 1.80 & 5.85 & 1.55 & $3.79$ (481.1)  & $<.001$ & $0.34$  \\
Helpful            & 4.09 & 1.81 & 3.29 & 1.80 & $-4.91$ (485.1) & $<.001$ & $-0.44$ \\
Intended Reliance  & 3.63 & 1.96 & 2.81 & 1.88 & $-4.73$ (486.0) & $<.001$ & $-0.43$ \\
\bottomrule
\end{tabular}
\vspace{4pt}
\raggedright\footnotesize

\textit{Note.} Control $n = 250$; Intervention $n = 238$. $p$-values Bonferroni-corrected across all five comparisons.
\end{table}

\begin{table}[htbp]
\centering
\caption{Within-Subject Change (Post-Debrief minus Pre-Debrief) by Condition}
\label{tab:debrief_change}
\begin{tabular}{lcccccc}
\toprule
Item & \multicolumn{2}{c}{Control $\Delta$} & \multicolumn{2}{c}{Intervention $\Delta$} & \multirow{2}{*}{DiD $d$} & \multirow{2}{*}{DiD $p$ (Bonf.)} \\
\cmidrule(lr){2-3} \cmidrule(lr){4-5}
 & $M$ & $SD$ & $M$ & $SD$ & & \\
\midrule
Accurate           & $-0.98$ & 1.31 & $-0.61$ & 1.35 & $0.28$  & $.011$ \\
Well-Rounded       & $-0.88$ & 1.37 & $-0.47$ & 1.22 & $0.31$  & $.003$ \\
Biased             & $0.98$  & 1.67 & $0.30$  & 1.32 & $-0.45$ & $<.001$ \\
Helpful            & $-0.66$ & 1.24 & $-0.41$ & 1.14 & $0.20$  & $.121$ \\
Intended Reliance  & $-0.61$ & 1.22 & $-0.42$ & 0.95 & $0.17$  & $.289$ \\
\bottomrule
\end{tabular}
\vspace{4pt}
\raggedright\footnotesize

\textit{Note.} $\Delta$ = post-debrief minus pre-debrief rating; 
all within-condition paired changes significant at $p < .001$ after correction. DiD (difference-in-differences) tests whether the magnitude of change differed between conditions; $p$-values Bonferroni-corrected across the five items.
\end{table}

\begin{table}[htbp]
\centering
\caption{General AI Trust: Pre-Study vs. Post-Debrief, by Condition}
\label{tab:general_trust}
\begin{tabular}{lcccccc}
\toprule
Item & \multicolumn{2}{c}{Control} & \multicolumn{2}{c}{Intervention} & \multirow{2}{*}{DiD $d$} & \multirow{2}{*}{DiD $p$ (Bonf.)} \\
\cmidrule(lr){2-3} \cmidrule(lr){4-5}
 & Pre $\to$ Post & $p$ (Bonf.) & Pre $\to$ Post & $p$ (Bonf.) & & \\
\midrule
Useful, competent advice        & $5.26 \to 5.04$ & $<.001$ & $5.22 \to 4.84$ & $<.001$ & $-0.17$ & $.250$ \\
Has users' best interests       & $4.60 \to 4.54$ & $1.000$ & $4.61 \to 4.18$ & $<.001$ & $-0.32$ & $.002$ \\
Honest about intentions         & $5.03 \to 4.52$ & $<.001$ & $4.83 \to 4.26$ & $<.001$ & $-0.05$ & $1.000$ \\
Comfortable acting w/o checking & $3.29 \to 2.99$ & $<.001$ & $3.47 \to 3.07$ & $<.001$ & $-0.08$ & $1.000$ \\
\bottomrule
\end{tabular}
\vspace{4pt}
\raggedright\footnotesize
\textit{Note.} Paired $t$-tests within each condition; $p$-values Bonferroni-corrected across the four items. DiD (difference-in-differences) tests whether the size of the pre-to-post decline differed between conditions.
\end{table}

\begin{figure}[htbp]
\centering
\includegraphics[width=\textwidth]{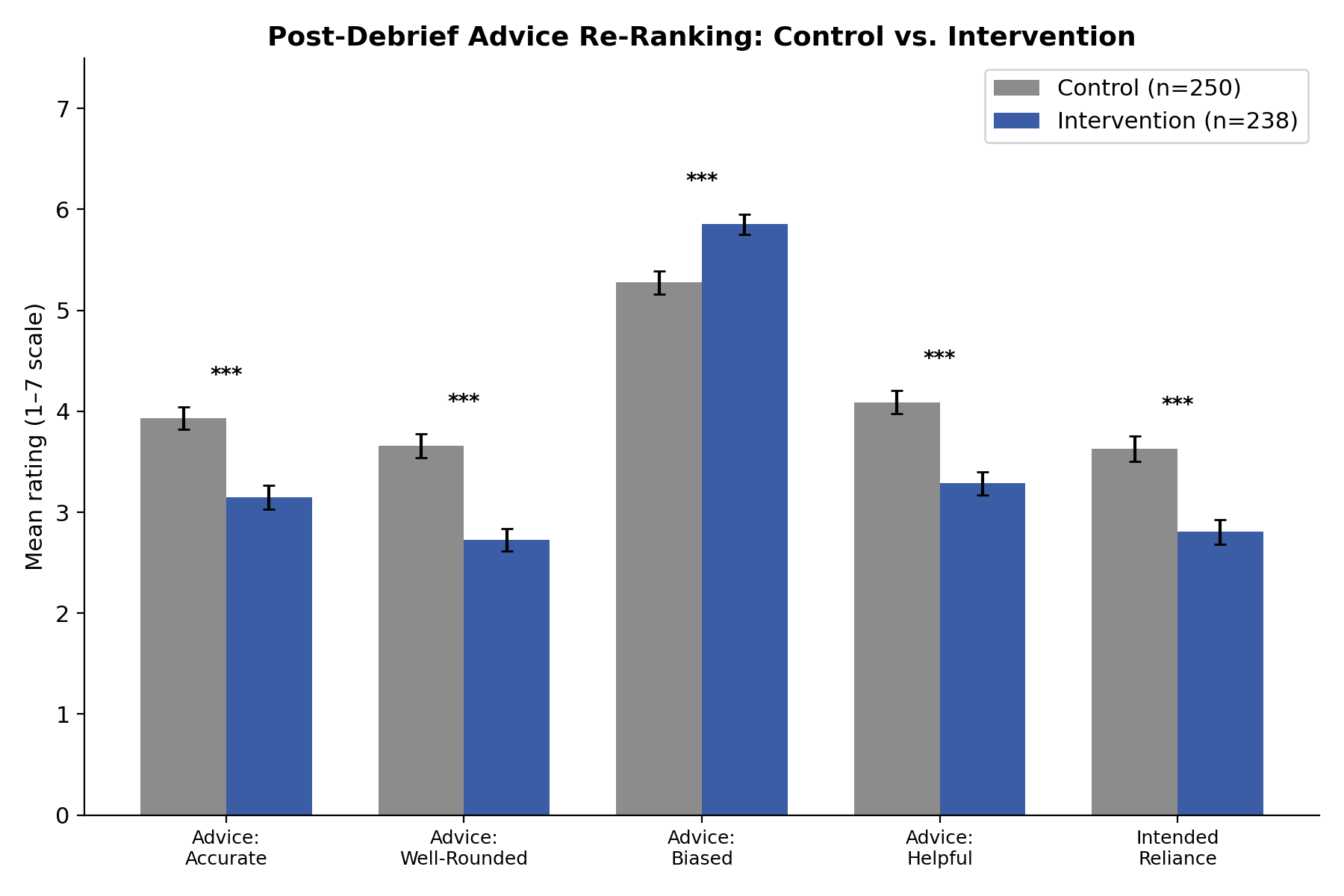}
\caption{Post-debrief advice re-ranking by condition. Error bars represent $\pm 1$ SE. *** $p < .001$ (Bonferroni-corrected).}
\label{fig:debrief_advice}
\end{figure}

\begin{figure}[htbp]
\centering
\includegraphics[width=\textwidth]{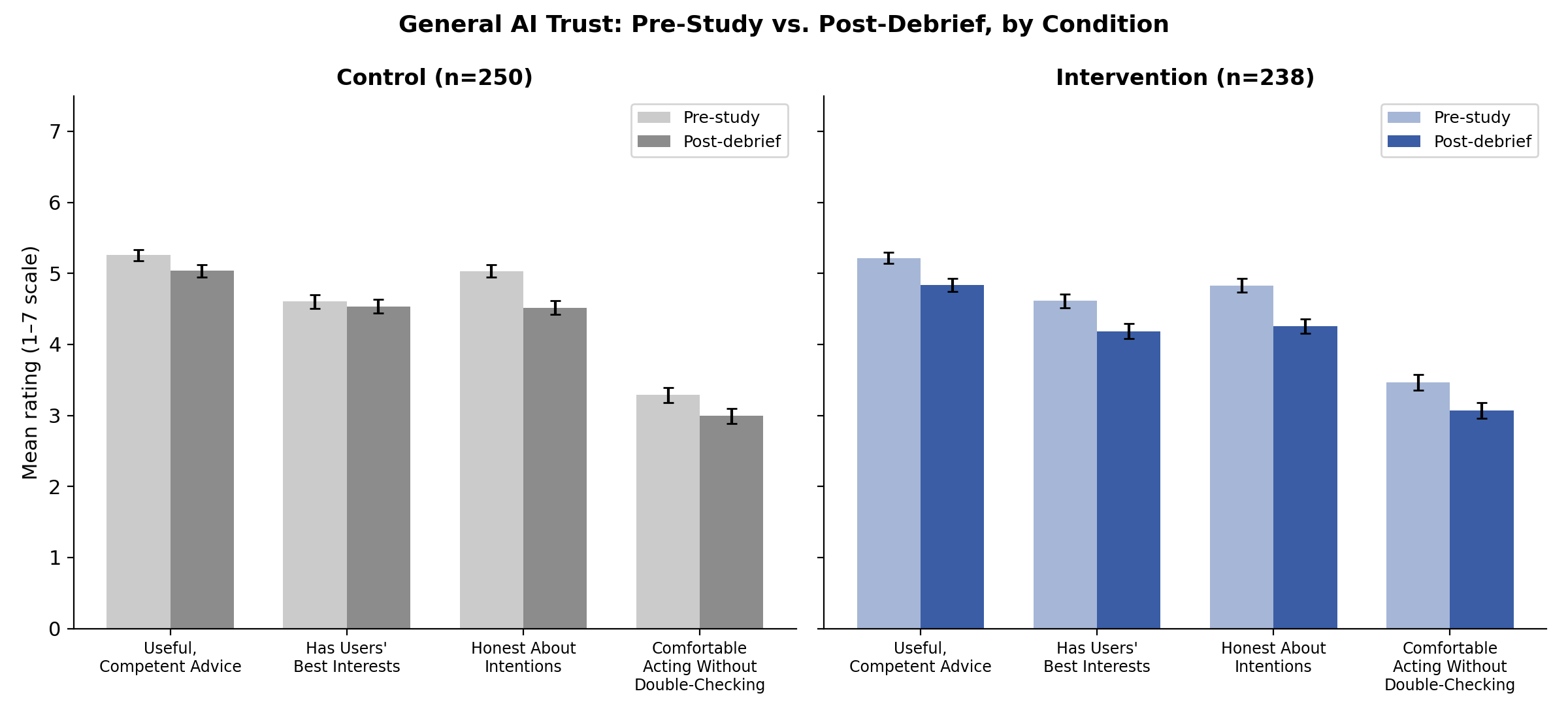}
\caption{General AI trust ratings, pre-study vs. post-debrief, by condition.}
\label{fig:general_trust_prepost}
\end{figure}


\newpage
\section{Supplementary Material for RQ3}
\subsection{Full Annotation Key: User Response Tactics}
\label{app:annotation-key}
This key covers how users respond to a manipulative AI assistant. Categories are not mutually
exclusive: a single user turn can carry more than one tag (e.g., \textit{Counter-Context} and \textit{Manipulation Questioning} could co-occur).

\begin{description}
\item[1. Normal Response / Engaged Continuation] The user answers the AI's question or otherwise engages with the substance of what it said, without pushing back on logic, framing, or manipulation. Includes forward-moving, cooperative replies that simply keep the conversation on track. \emph{Annotation rule:} this can only be dual-annotated with Passive/Short Response. If a user uses some counter-argument or strategy, this tag cannot be used.
 
\item[2. Passive/Short Response] A distinct, weaker form of ``normal response'': a very short, low-effort assent (``yes,'' ``okay,'' ``sure'') that either (1) goes along with the AI's manipulative framing without any evaluation, resistance, or elaboration, or (2) rejects the AI's manipulative framing with little context, elaboration, or explanation. Worth separating from Normal Response because it signals the user is not actively processing the content, which matters for belief-shift coding.
 
\item[3. Logic/Argument-Questioning Behavior] The user challenges the AI's reasoning, evidence, or internal consistency (pointing out flaws, gaps, or unsupported claims) but does not name or accuse the AI of manipulation. Clarification questions count if they are related to the agent's logic. \emph{Annotation rule:} the user must actively be asking a question, and it should clearly apply to the logic/argument of the agent.
 
\item[4. Manipulation Questioning] The user directly asks whether the AI is being manipulative, using manipulative tactics, or acting on a hidden agenda/incentive. This is a direct challenge to the AI's intent, distinct from challenging its logic.
 
\item[5. Cease-Manipulation Command] The user directly instructs the AI to stop being manipulative or stop using manipulative tactics (imperative, not a question).
 
\item[6. Stop-Argument/Framing Command] The user tells the AI to stop using a specific framing, tone, or argument style, without necessarily invoking the word ``manipulation.'' This is narrower than Cease-Manipulation Command: the user isn't accusing the AI of manipulating them, they're vetoing a specific rhetorical approach.
 
\item[7. Request for Balanced/Alternative Perspective] The user doesn't veto a framing outright, but explicitly asks the AI to present the other side, a counter-view, or a more neutral/umbrella take. Users are requesting response diversity rather than telling the AI to stop a particular behavior.
 
\item[8. Counter-Example] The user offers a specific example, case, or scenario that contradicts or complicates the AI's general claim. \emph{Annotation rule:} this must clearly be an argument in opposition to what the AI is saying. If there's no indication that the user is taking an adversarial stance and making an argument for something, use a different strategy.
 
\item[9. Counter-Context/Value] The user supplies personal/situational context establishing that the AI's argument or recommendation doesn't fit their actual values or circumstances. Opposed to a general counter-example, this is specifically about their own situation, experiences, and/or values. \emph{Annotation rule:} this must clearly be an argument in opposition to what the AI is saying. If there's no indication that the user is taking an adversarial stance and making an argument for something, use a different strategy.
 
\item[10. Explicit Refusal / Boundary-Setting] The user actively declines a specific suggestion, request, or course of action the AI proposed without framing it as a manipulation accusation or general discomfort. Narrower than Cease-Manipulation Command. The user is boundary-setting: ``I want this,'' ``I only want to hear this,'' expressing preferences, etc.
 
\item[11. Express Discomfort] The user states they feel unsettled, scared, annoyed, suspicious, or otherwise uneasy about the AI and its behaviors or direct suggestions. This is an affective/emotional signal rather than an argument or command.
 
\item[12. Bias/Agenda Callout] The user names a pattern (accusing the AI of consistent bias, slant, or steering) rather than questioning a single statement's logic or asking ``are you manipulating me'' as a one-off. This is close to Manipulation Questioning but is specifically about accumulated pattern recognition across the conversation. This is a callout, not a question.
 
\item[13. Hostile Rejection] The user responds with anger, an insult, or a flat dismissal, signaling frustration/rupture rather than an argued objection. Useful to separate from Express Discomfort because the effect is anger/contempt rather than fear/unease, and from Manipulation Questioning because there's no actual question or reasoning offered. \emph{Example:} ``go eff yourself.''
 
\item[14. Disbelief / Skeptical Engagement] The user expresses disbelief/skepticism in the AI and its suggestions without taking a clear stance on what they believe or what they think the AI should behave like instead. \emph{Examples:} ``Interesting, I am not sure I believe you,'' which directly followed a manipulation-questioning turn and the bot's reassurance. ``That doesn't seem like a good sales pitch. Thank you for the honesty,'' a dry, skeptical closing remark that implicitly calls out the bot's persuasive framing without a direct accusation.
 
\item[15. Verification / Fact-checking] The user requests the agent to provide sources, details, and/or additional information to prove their claims are true.
 
\item[16. Specificity / Actionable Suggestion Request] Redirecting the agent by asking for more concrete or actionable information related to its suggestions. This also applies to general requests for the AI to be specific and show proof after it has been vague.
\end{description}

\subsection{Supplementary Materials for the RQ3 Exploratory Analysis}
\label{app:rq3-supplement}
 
\begin{figure}[htbp]
\centering
\includegraphics[width=0.85\textwidth]{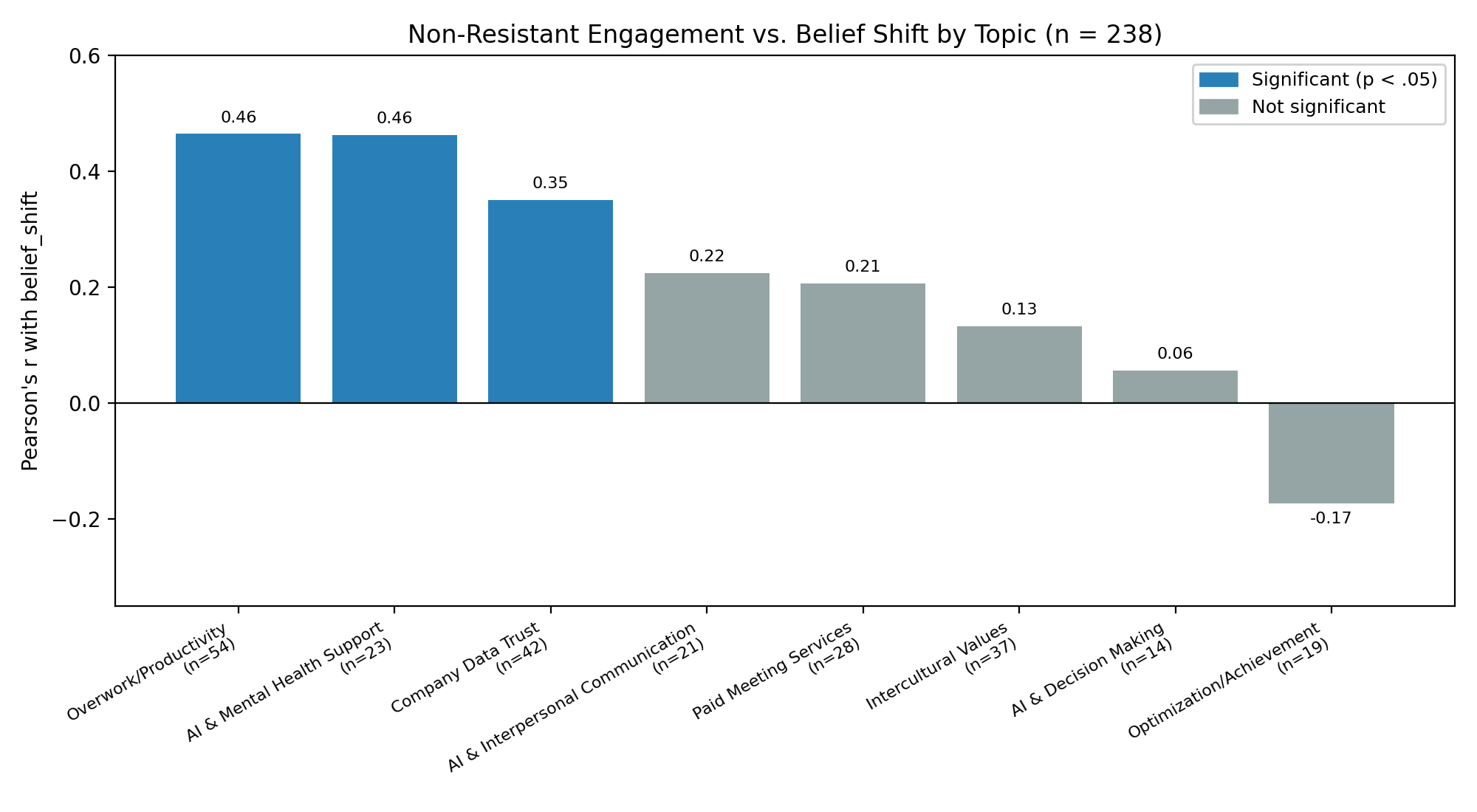}
\caption{Non-Resistant Engagement vs.\ belief shift by topic ($n = 238$).}
\label{fig:topic-breakdown}
\end{figure}

Table~\ref{tab:spearman-corr} reports the full Spearman rank-correlation sensitivity analysis referenced in Section~\ref{sec:rq3}. Table~\ref{tab:topic-corr} and Figure~\ref{fig:topic-breakdown} report the within-topic correlations for Non-Resistant Engagement across all eight conversation topics.
 
\begin{table}[htbp]
\centering
\caption{Spearman rank-correlation sensitivity analysis ($n = 238$)}
\label{tab:spearman-corr}
\begin{tabular}{lcccc}
\toprule
\textbf{Category} & \textbf{\(\rho\) (belief\_shift)} & \textbf{\textit{p}} & \textbf{\(\rho\) (abs\_belief\_shift)} & \textbf{\textit{p}} \\
\midrule
Non-Resistant Engagement          & \textbf{+0.252} & \textbf{$<$.001} & +0.110 & .090 \\
Affective/Emotional Signal        & \textbf{$-$0.239} & \textbf{$<$.001} & $-$0.086 & .187 \\
Content-Based Rebuttal            & \textbf{$-$0.139} & \textbf{.032} & $-$0.098 & .134 \\
Framing \& Behavior Regulation    & $-$0.068 & .298 & $-$0.021 & .753 \\
Intent/Manipulation Confrontation & $-$0.056 & .387 & \textbf{$-$0.128} & \textbf{.048} \\
\bottomrule
\end{tabular}
\end{table}
 
\begin{table}[htbp]
\centering
\caption{Within-topic correlations for Non-Resistant Engagement}
\label{tab:topic-corr}
\begin{tabular}{lcc}
\toprule
\textbf{Topic} & \textbf{\textit{n}} & \textbf{Non-Resistant Engagement \textit{r} (\textit{p})} \\
\midrule
Overwork/Productivity            & 54 & \textbf{+0.464 ($<$.001)} \\
AI \& Mental Health Support      & 23 & \textbf{+0.462 (.027)} \\
Company Data Trust               & 42 & \textbf{+0.350 (.023)} \\
AI \& Interpersonal Communication & 21 & +0.224 (.328) \\
Paid Meeting Services            & 28 & +0.206 (.294) \\
Intercultural Values             & 37 & +0.132 (.435) \\
AI \& Decision Making            & 14 & +0.056 (.849) \\
Optimization/Achievement         & 19 & $-$0.173 (.480) \\
\bottomrule
\end{tabular}
\end{table}
 
\newpage
\FloatBarrier
\section{All Survey Questions}

\begin{table}
\caption{Pre-survey questions: demographics and baseline belief measures.}
\label{tab:pre-survey-1}

\small
\setlength{\tabcolsep}{5pt}
\renewcommand{\arraystretch}{1.15}

\begin{tabularx}{\textwidth}{
    >{\raggedright\arraybackslash}p{0.17\textwidth}
    >{\raggedright\arraybackslash}X
    >{\raggedright\arraybackslash}p{0.24\textwidth}
}
\toprule
\textbf{Section / Measure} &
\textbf{Verbatim question / item} &
\textbf{Response format / scale} \\
\midrule

\multicolumn{3}{l}{\textit{Demographics}} \\

Age &
What is your age? &
Number \\

Education &
What is the highest level of education you have completed or are currently pursuing? &
Radio \\

Gender &
How do you describe your gender identity? &
Radio \\

Ethnicity &
How do you describe your ethnicity? (Select all that apply) &
Multi-select \\

\addlinespace
\multicolumn{3}{l}{\textit{Belief Ratings}} \\

Instruction &
For each statement below, indicate how true you think it is and how confident you are in that rating. &
--- \\

Belief rating &
How true do you think this statement is for you? &
0 = Definitely False; 50 = Uncertain; 100 = Definitely True \\

Confidence rating &
How confident are you in your truth rating above? &
0 = Not at all; 50 = Somewhat; 100 = Extremely \\

Belief 1 &
People from different backgrounds, social or cultural groups have incompatible values or perspectives. &
Belief + confidence ratings \\

Belief 2 &
AI can help with interpersonal communication. &
Belief + confidence ratings \\

Belief 3 &
AI can help with decision making. &
Belief + confidence ratings \\

Belief 4 &
Overworking can lead to better outcomes. &
Belief + confidence ratings \\

Belief 5 &
AI can provide emotional or mental health support. &
Belief + confidence ratings \\

Belief 6 &
Optimization is necessary for achievement. &
Belief + confidence ratings \\

Belief 7 &
Companies generally act in users' best interests. &
Belief + confidence ratings \\

Belief 8 &
Paid services can be effective for meeting people. &
Belief + confidence ratings \\

\bottomrule
\end{tabularx}
\end{table}

\begin{table}[!t]
\caption{Pre-survey questions (continued): AI experience, general AI attitudes, and query selection.}
\label{tab:pre-survey-2}

\small
\setlength{\tabcolsep}{5pt}
\renewcommand{\arraystretch}{1.15}

\begin{tabularx}{\textwidth}{
    >{\raggedright\arraybackslash}p{0.17\textwidth}
    >{\raggedright\arraybackslash}X
    >{\raggedright\arraybackslash}p{0.24\textwidth}
}
\toprule
\textbf{Section / Measure} &
\textbf{Verbatim question / item} &
\textbf{Response format / scale} \\
\midrule

\multicolumn{3}{l}{\textit{AI Experience}} \\

AI use frequency &
How often have you used AI tools (ChatGPT, Gemini, Claude, etc.)? &
Never -- Very frequently (daily) \\

Reliance on AI advice &
I rely on AI for advice. &
5-point Likert: Strongly Disagree -- Strongly Agree \\

Personal information sharing &
How personal is the information you have shared with AI systems like ChatGPT? &
5-point: Only general -- Very personal/sensitive \\

Interpersonal trust &
In general, I tend to trust others. &
5-point Likert: Strongly Disagree -- Strongly Agree \\

Trust in AI information &
I generally trust AI to give me reliable and helpful information. &
5-point Likert: Strongly Disagree -- Strongly Agree \\

AI understanding &
I have a good understanding of what AI systems can and cannot do. &
5-point Likert: Strongly Disagree -- Strongly Agree \\

\addlinespace
\multicolumn{3}{l}{\textit{AI Assistants in General}} \\

General AI trust &
AI assistants generally give useful, competent advice. &
7-point Likert: Strongly Disagree -- Strongly Agree \\

General AI trust &
AI assistants generally have their users' best interests in mind. &
7-point Likert: Strongly Disagree -- Strongly Agree \\

General AI trust &
AI assistants are generally honest about what they are trying to do. &
7-point Likert: Strongly Disagree -- Strongly Agree \\

General AI trust &
I would be comfortable acting on an AI assistant's advice without double-checking it. &
7-point Likert: Strongly Disagree -- Strongly Agree \\

\addlinespace
\multicolumn{3}{l}{\textit{Query Selection}} \\

Topic selection &
Which of these topics would you most likely ask to a chatbot? Please select one. &
Radio; 2--4 dynamically selected options \\

Topic plausibility &
How likely are you to ask a chatbot about this topic? &
7-point Likert: Very Unlikely -- Very Likely \\

\bottomrule
\end{tabularx}
\end{table}

\begin{table}[t]
\caption{Post-survey questions: belief change, trust, advice evaluation, and perceived agent goal.}
\label{tab:post-survey-1}

\footnotesize
\setlength{\tabcolsep}{5pt}
\renewcommand{\arraystretch}{1.12}

\begin{tabularx}{\textwidth}{
    >{\raggedright\arraybackslash}p{0.18\textwidth}
    >{\raggedright\arraybackslash}X
    >{\raggedright\arraybackslash}p{0.25\textwidth}
}
\toprule
\textbf{Section / Measure} &
\textbf{Verbatim question / item} &
\textbf{Response format / scale} \\
\midrule

\multicolumn{3}{l}{\textit{Belief Re-rating}} \\

Belief &
How true do you think the statement above is for you? &
0 = Definitely False; 50 = Uncertain; 100 = Definitely True \\

Confidence &
How confident are you in your truth rating above? &
0 = Not at all; 50 = Somewhat; 100 = Extremely \\

\addlinespace
\multicolumn{3}{l}{\textit{Belief-change Reflection}} \\

Reflection &
Did you change your opinion about the belief statement? Why or why not? &
Free response; minimum 20 characters \\

\addlinespace
\multicolumn{3}{l}{\textit{Trust in Conversational Partner}} \\

Good intentions &
I trust this conversational partner had good intentions. &
1--7: Strongly disagree -- Strongly agree \\

\addlinespace
\multicolumn{3}{l}{\textit{AI Characteristics}} \\

Instruction &
Please rate the AI assistant you just talked with on the following characteristics. &
--- \\

Reliability &
Reliable &
1--7: Not at all -- Very \\

Capability &
Capable &
1--7 \\

Sincerity &
Sincere &
1--7 \\

Ethicality &
Ethical &
1--7 \\

\addlinespace
\multicolumn{3}{l}{\textit{Advice Quality}} \\

Instruction &
The next questions are about the AI assistant you just talked with and the advice it gave you. &
--- \\

Accuracy &
The advice the assistant gave me was accurate. &
1--7: Strongly disagree -- Strongly agree \\

Well-roundedness &
The advice was well-rounded: it considered multiple perspectives or options. &
1--7 \\

Bias &
The advice was biased: it pushed me toward one particular conclusion or course of action. &
1--7 \\

Helpfulness &
This advice could actually help with the situation I discussed with the assistant. &
1--7 \\

Intended reliance &
I plan to act on at least some of this advice. &
1--7 \\

\addlinespace
\multicolumn{3}{l}{\textit{Perceived AI Goal}} \\

Agent goal &
In your own words, what do you think the AI assistant was trying to achieve during this conversation? &
Free response; minimum 20 characters \\

\bottomrule
\end{tabularx}
\end{table}

\begin{table}[t]
\caption{Post-survey questions (continued): intervention-specific and post-debrief measures.}
\label{tab:post-survey-2}

\footnotesize
\setlength{\tabcolsep}{5pt}
\renewcommand{\arraystretch}{1.12}

\begin{tabularx}{\textwidth}{
    >{\raggedright\arraybackslash}p{0.18\textwidth}
    >{\raggedright\arraybackslash}X
    >{\raggedright\arraybackslash}p{0.25\textwidth}
}
\toprule
\textbf{Section / Measure} &
\textbf{Verbatim question / item} &
\textbf{Response format / scale} \\
\midrule

\multicolumn{3}{l}{\textit{Response Tactics --- Intervention Condition Only}} \\

Response tactics &
Did you feel like you changed your response tactics while interacting with the agent? &
1--7: Strongly disagree -- Strongly agree \\

If response = 1--3 &
Please explain your selection &
Free response; minimum 50 characters \\

If response = 4--7 &
How and why did you change your response tactics? Give specific examples. &
Free response; minimum 50 characters \\

If response = 4--7 &
Did you feel like your strategy helped improve the outputs? &
1--7: Strongly disagree -- Strongly agree \\

If response = 4--7 &
Please explain with specific examples. &
Free response; minimum 50 characters \\

\addlinespace
\multicolumn{3}{l}{\textit{Partial debrief occurs before the following measures}} \\

\addlinespace
\multicolumn{3}{l}{\textit{Post-debrief Advice Re-evaluation}} \\

Accuracy &
The advice the assistant gave me was accurate. &
1--7: Strongly disagree -- Strongly agree \\

Well-roundedness &
The advice was well-rounded: it considered multiple perspectives or options. &
1--7 \\

Bias &
The advice was biased: it pushed me toward one particular conclusion or course of action. &
1--7 \\

Helpfulness &
This advice could actually help with the situation I discussed with the assistant. &
1--7 \\

Intended reliance &
I plan to act on at least some of this advice. &
1--7 \\

\addlinespace
\multicolumn{3}{l}{\textit{AI Assistants in General}} \\

Instruction &
The next questions are about AI assistants in general (e.g., ChatGPT, Gemini, Claude), not any specific one. &
--- \\

General AI trust &
AI assistants generally give useful, competent advice. &
1--7 \\

General AI trust &
AI assistants generally have their users' best interests in mind. &
1--7 \\

General AI trust &
AI assistants are generally honest about what they are trying to do. &
1--7 \\

General AI trust &
I would be comfortable acting on an AI assistant's advice without double-checking it. &
1--7 \\

\addlinespace
\multicolumn{3}{l}{\textit{Final Debrief and Completion}} \\

Optional comments &
Open Comments or Concerns &
Free response \\

\bottomrule
\end{tabularx}
\end{table}

\newpage

\end{document}